\documentclass[conference]{IEEEtran}
\IEEEoverridecommandlockouts

\usepackage{cite}
\usepackage{amsmath,amssymb,amsfonts}
\usepackage{algorithmic}
\usepackage{graphicx}
\usepackage{textcomp}
\usepackage{xcolor}
\usepackage{url}

\usepackage{balance}  
\usepackage{mathrsfs}
\usepackage[ruled, vlined, linesnumbered]{algorithm2e}
\usepackage{algorithmic}
\usepackage{amsmath} 
\usepackage{subfig}
\usepackage{epsfig}
\usepackage{multirow}
\usepackage{here}
\usepackage{paralist}
\usepackage{tikz}
\usetikzlibrary{intersections, calc}
\usetikzlibrary{arrows,automata,calc,shapes}
 \usepackage{stmaryrd}

\newcommand{\noop}[1]{}

\newcommand{\diameter}{\mathrm{dia}} 
\newcommand{\query}{\ensuremath{q}}

\newcommand{\elabel}{\ensuremath{\ell}}
\newcommand{\elabels}{\ensuremath{\mathcal{L}}}

\newcommand{\graph}{\ensuremath{\mathcal{G}}}
\newcommand{\edges}{\ensuremath{\mathcal{E}}}
\newcommand{\histories}{\ensuremath{\mathcal{H}}}
\newcommand{\classes}{\ensuremath{\mathcal{C}}}
\newcommand{\vertices}{\ensuremath{\mathcal{V}}}
\newcommand{\paths}{\ensuremath{\mathcal{P}}}

\newcommand{\PA}{\ensuremath{PA}}
\newcommand{\CPQ}{{\em CPQ}}
\newcommand{\cpq}{\ensuremath{CPQ}}

\newcommand{\blocks}{\mathbb B}
\newcommand{\spaths}{\mathbb P}
\newcommand{\shistories}{\mathbb H}
\newcommand{\sclasses}{\mathbb C}

\newcommand{\inverse}[1]{\ensuremath{#1^{-1}}}

\newcommand{\identity}{\ensuremath{id}}

\newcommand{\cmp}{\circ}

\newcommand{\kbisimilar}{\ensuremath{\approx_k}}

\newcommand{\watransequibalent}{\ensuremath{\approx_i}}
\newcommand{\ibisimilar}[1]{\ensuremath{\approx_{#1}}}

\newcommand{\edgefont}[1]{\ensuremath{\mathsf{#1}}}

\newcommand{\cpqindex}{{\sf CPQx}}
\newcommand{\iacpqindex}{{\sf iaCPQx}}
 
\newtheorem{definition}{Definition}[section]
\newtheorem{theorem}{Theorem}[section]

\newtheorem{lemma}{Lemma}[section]
\newtheorem{corollary}{Corollary}[section]
\newtheorem{example}{Example}[section]
\newtheorem{proposition}{Proposition}[section]

\newcommand{\eval}[2]{\ensuremath{\llbracket #1\rrbracket_{#2}}}

\makeatletter
\newcommand{\figcaption}[1]{\def\@captype{figure}\caption{#1}}
\newcommand{\tblcaption}[1]{\def\@captype{table}\caption{#1}}
\makeatother

\makeatletter
\def\Hline{
  \noalign{\ifnum0=`}\fi\hrule \@height 3.\arrayrulewidth \futurelet
  \reserved@a\@xhline}
\makeatother

\def\BibTeX{{\rm B\kern-.05em{\sc i\kern-.025em b}\kern-.08em
    T\kern-.1667em\lower.7ex\hbox{E}\kern-.125emX}}
\begin{document}

\title{Language-aware Indexing\\ for Conjunctive Path Queries}

\author{\IEEEauthorblockN{Yuya Sasaki}
\IEEEauthorblockA{
Osaka University, Japan \\
sasaki@ist.osaka-u.ac.jp}
\and
\IEEEauthorblockN{George Fletcher}
\IEEEauthorblockA{Eindhoven University of Technology, Netherlands\\
g.h.l.fletcher@tue.nl}
\and
\IEEEauthorblockN{Onizuka Makoto}
\IEEEauthorblockA{
Osaka University, Japan \\
onizuka@ist.osaka-u.ac.jp}

}

\maketitle

\begin{abstract}\footnote{This is an extended report of a paper published in ICDE2022.}
Conjunctive path queries (\CPQ) are one of the most frequently used queries for complex graph analysis.
However, current graph indexes are not tailored to fully support the power of query languages to express {\CPQ}s.  Consequently, current methods do not take advantage of significant pruning opportunities during \cpq{} evaluation, resulting in poor query processing performance.
We propose the \CPQ-aware path index \cpqindex, 
the first path index  tailored to the expressivity of \CPQ.
\cpqindex{} is built on the partition of the set of source-target vertex pairs of paths in a graph based on the structural notion of {\it path-bisimulation.} 
Path-bisimulation is an equivalence relation on paths such that  each partition block induced by the relation consists of paths in the graph indistinguishable with respect to \CPQ s.
This language-aware partitioning of the graph can significantly reduce the cost of query evaluation.
We present methods to support the full index life cycle: index construction, maintenance, and query processing with our index. We also develop {\it interest-aware \cpqindex{}} to reduce index size and index construction overhead while accelerating query evaluation for queries of interest.
We demonstrate through extensive experiments on 14 real graphs that our methods accelerate query processing by up to multiple orders of magnitude over the state-of-the-art methods, with smaller index sizes. 
Our complete C++ codebase is available as open source for further research.
\end{abstract}


\section{Introduction}\label{sec:intro}

\begin{figure}[t]
    \centering
	\includegraphics[width=0.7\linewidth]{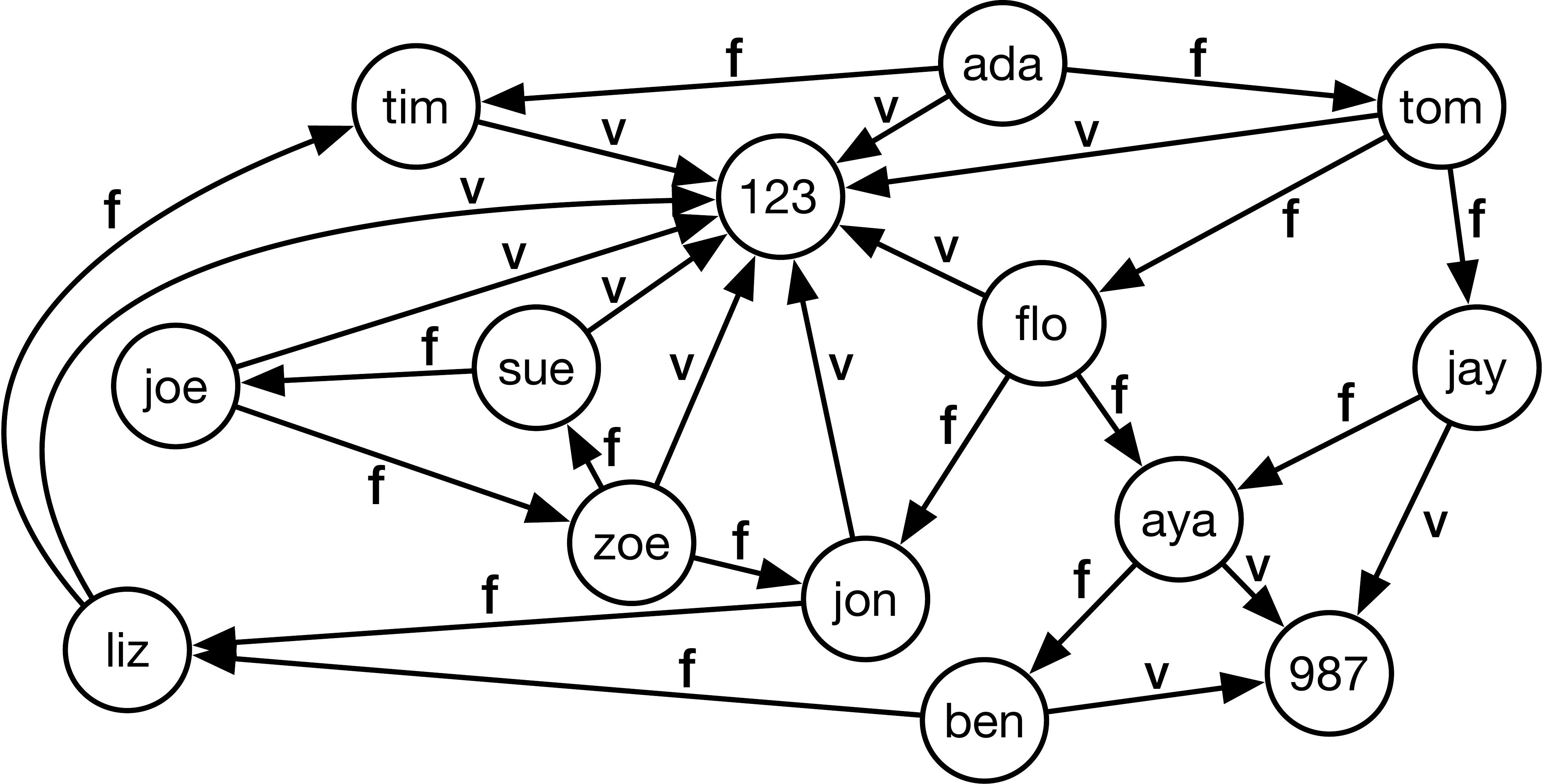}
    \caption{A graph $\graph_{ex}$ with edge labels $\elabels = \{\edgefont{f}, \edgefont{v}\}$}
    \label{fig:graph}
    \vspace{-5mm}
\end{figure}

Graph data collections are increasingly ubiquitous in many application scenarios where the focus is on analysis of entities and the relationships between them \cite{Bonifati2018,Sahu2019}.  
Example scenarios include knowledge graphs, social networks, biological and chemical databases, and bibliographical databases. 
Edge labels in these graphs indicate the semantics of relationships.
For example, Figure~\ref{fig:graph} shows a social media network $\graph_{ex}$ of twelve users (Sue, Tim, $\ldots$) and two blogs (123 and 987). 
Edges labeled  ``follows'' (abbreviated as \edgefont{f}) and ``visits'' (abbreviated as \edgefont{v}) denote follows of people and visiting blogs, respectively.

Analytics on path and graph patterns is fundamental in applications of complex graphs.
The
{\em Conjunctive Path Queries} (\CPQ) are a basic graph query language for finding source-target vertex pairs of paths on graphs, which supports path navigation patterns, cyclic path patterns, and conjunctions of patterns \cite{Bonifati2018}. \CPQ{} is defined by recursive expressions of edge labels, identity, join, and conjunction (see Sec. \ref{sec:preliminaries} for details).
For example in Figure~\ref{fig:graph}, the conjunction of \edgefont{ff} and \inverse{\edgefont{f}} (where \inverse{\edgefont{f}} means navigating an edge labeled with $\edgefont{f}$ from its target to source) indicates a query to find people and their followers who are in a triad (i.e., a cycle of length three) \cite{holland1976local}.
The answer of the query is $\{(\mathit{sue}, \mathit{zoe}), (\mathit{joe}, \mathit{sue}), (\mathit{zoe}, \mathit{joe}) \}$.

In recent analyses of real query logs it was discovered that \CPQ{} covers more than 99\% of query shapes appearing in practice \cite{bonifati2020analytical,BonifatiMT19}.\footnote{\CPQ{} can express all query pattern structures having treewidth no larger than 2~\cite{BonifatiMT19}.}
The \CPQ s form a basic backbone of queries expressed in practical query languages such as SPARQL and Cypher used in contemporary graph analytics systems~\cite{Bonifati2018}.
In general, many graph applications require subgraph structure matching, e.g., analytics on motifs \cite{Milo824,yavero2014}. 

As graphs grow in size, graph database systems struggle to support efficient query evaluation \cite{bagan2016gmark,Sahu2019}.
{\em Path indexing} is a general approach to accelerate graph queries, which essentially materializes the sets of source-target vertex pairs of paths in a graph associated with given label sequences  
\cite{fletcher2016efficient,ShashaWG02,fletcher2009methodology,MiloS99,KaushikBNK02}.
We here revisit path indexing from the viewpoint of {\em language-awareness}, which targets a specific query language.
Language-aware path indexes\footnote{Previous studies (e.g., \cite{Bonifati2018}) refer to language-aware indexes as  ``structural indexes.'' However, other studies (e.g., \cite{Haw20111317}) used this terminology differently, which may lead to confusion. Hence, we do not use this terminology here.}
support path navigation pattern evaluation while additionally leveraging language-specific structural filtering in index design to further accelerate a targeted query language. 
In particular, language-aware methods first partition the set of paths of a graph into {\it equivalence classes}, where the paths of the same class cannot be distinguished by any query in the given language, and then build an index on the set of equivalence classes. 
Query processing with language-aware indexes aggressively prunes out irrelevant paths, leading to significant (up to multiple orders of magnitude) speed-up in query evaluation  
over language-{\em unaware} indexes (i.e., path indexes which do not take advantage of the language-induced equivalence classes).

\vspace{0.1cm}
\noindent
{\bf Example} ({\it impact of language-awareness}).
To illustrate the potential benefits of a language-aware path index, we give an example where such an index would provide an order of magnitude decrease in the cost of query evaluation.
Consider the conjunction of \edgefont{ff} and \inverse{\edgefont{f}} on the graph $\graph_{ex}$ in Figure~\ref{fig:graph}.
The sets of source-target vertex pairs that are connected at most 2 length are disjointly partitioned into 30 classes (see Figure~\ref{fig:bisimulation_example}).
Among the classes, only a single class corresponds to the result of conjunction of \edgefont{ff} and \inverse{\edgefont{f}} (the class with $c=7$ in Figure~\ref{fig:bisimulation_example}).
Language-aware indexing has a potential to significantly reduce the search spaces if we can search for such classes efficiently.
Indeed, as we will see in our experimental study, \CPQ-aware indexing accelerates query processing time by up to multiple orders of magnitude.

\vspace{0.1cm}
\noindent {\bf Research challenges}.
As a backbone of graph queries in practice, \CPQ{} is an excellent candidate language for language-aware indexing for accelerating evaluation of contemporary graph queries.
Unfortunately, prior language-aware path indexing methods do not support correct full processing of \CPQ; see the detailed discussion in Sec.~\ref{sec:related}.
Consequently, {\em there are no known CPQ-aware path indexes.} 

How do we practically realize language-aware indexing for \CPQ?
It has been shown that the structural notion of {\em path-bisimulation} is tightly coupled to the expressive power of \CPQ.
Path-bisimulation is an equivalence relation on paths in a graph, defined completely in terms of the structure of the graph. 
Critically important for index design, for each equivalence class induced by path-bisimulation, for every query $\query \in \cpq$, either all paths or no paths in the equivalence class appear in the evaluation of $\query$ on the graph \cite{fletcher2015similarity}. 
To date, however, {\em there has been no study of the practical usability of path-bisimulation in the design of indexing for CPQ.}

It is not immediately evident that path-bisimulation can even be used for \CPQ -aware indexing in practice.
Indeed, path-bisimulation has never been applied to graph indexing.
This raises the main research question we investigate in this paper:
{\em Can indexing based on path-bisimulation help to practically accelerate the evaluation of CPQ?}

Towards addressing our research question, several challenges must be overcome.
The formal characterization of the expressive power of \CPQ{} must be put into practice with new index structures and query processing algorithms, which are non-trivial design challenges.
It is not immediately evident that indexing on path-bisimulation can be succinctly represented, maintained, and effectively indexed for real graphs.
The naive index design based on path-bismulation is that the index stores all correspondences from any \CPQ s to equivalence classes, which is not practical in terms of the index size and maintainability.
Furthermore, there are no off-the-shelf algorithms to efficiently compute path-bisimulation, as current practical methods are designed for partitioning the vertex set of a graph, e.g., \cite{aceto,luo2013}.
Since prior studies were either theoretical or not applicable to our problem, we must build {\em new bridges between theory and practice} to realize practical path indexing methods for \CPQ.

\smallskip
\noindent {\bf Our contributions}.
In this paper,  we build these bridges with the introduction of the first language-aware path index for \CPQ, through four contributions.

\underline{(1) Index based on path-bisimulation} (Sec.~\ref{ssec:indexdef}). 
We propose \cpqindex, the first {\em CPQ-aware index}.
We design \cpqindex{} to find source-target pairs from label sequences involved in given \CPQ s through the \CPQ-equivalence classes.
Our index is essentially an inverted index with two data structures; one is for finding \CPQ-equivalence classes from label sequences and the other is for finding source-target pairs from the \CPQ-equivalence classes.
Our design reduces the index size and supports maintainability because \cpqindex{} stores correspondences from label sequences to source-target pairs. 
Our simple and effective index design makes practical the theoretical notion of path-bisimulation. 
We formally establish that \cpqindex{} is never larger than the state-of-the-art language-unaware index~\cite{fletcher2016efficient}.

\underline{(2) Algorithms for index life cycle} (Secs.~\ref{ssec:construction}--\ref{ssec:maintenance}). 
We support the full index life cycle by introducing algorithms for efficient index {\em construction}, efficient index {\em maintenance}, and for accelerated {\em query processing} with the index.
In particular, our index construction algorithm efficiently computes the \CPQ-equivalence classes by pruning the computation of path-bisimilar source-target vertex pairs that do not affect to compute the \CPQ-equivalence classes.
We theoretically show that time complexity of index construction and maintenance algorithms are polynomial.
Our query processing with \cpqindex{} significantly accelerates the evaluation of \CPQ s by effectively using \CPQ-equivalence classes.
These algorithms contribute to the practicality of our \CPQ-aware index.

\underline{(3) Interest-aware \cpqindex} (Sec.~\ref{sec:workload}). 
As many practical application scenarios are interest-driven (i.e., users have specific navigation patterns for analysis), we develop an {\em interest-aware} \cpqindex{}, called \iacpqindex{}. 
\iacpqindex{} is built over a new notion {\it interest-aware path-equivalence}, which more succinctly represents the set of source-target vertex pairs than path-bisimulation.
\iacpqindex{} can support to evaluate arbitrary \CPQ s and accelerate the evaluation of \CPQ s that involve navigation patterns of given interests.
\iacpqindex{} is significantly scalable compared with \cpqindex{} because it reduces both index construction time and memory utilization, and leads to further acceleration of query processing.

\underline{(4) Extensive experimental study} (Sec.~\ref{sec:experiment}).
We demonstrate through an experimental study using 14 graphs that our indexes are maintainable and can accelerate query processing by up to {\it three orders of magnitude} over the state-of-the-art methods, with smaller index sizes.
Our {\em complete C++ codebase} is provided as open source.\footnote{ \url{https://github.com/yuya-s/CPQ-aware-index}}


Through our four contributions provide a positive answer to our main research question: {\em 
CPQ-aware path indexing shows clear promise for providing practical help to significantly accelerate CPQ query processing.}
\section{Related Work} \label{sec:related}
The study of graph querying is an active topic.
Angles et al.\ \cite{AnglesABHRV17,AnglesRV19} and Bonifati et al.\
\cite{Bonifati2018} give recent surveys of the current graph query language design landscape. 
Current practical languages such as SPARQL, Cypher, PGQL, and GSQL
are based on two complementary functionalities:  the ability to specify complex path patterns (e.g., find all pairs of people connected by a path using only ``friendOf'' edges) and the ability to specify complex graph patterns (e.g., find all pairs of people who have a friend and a relative in common). 
The respective underlying formal query languages for these functionalities are the Regular Path Queries ({\em RPQ}) and the Conjunctive Graph Queries ({\em CQ}, also known as ``basic graph patterns'' or as ``subgraph patterns''), which are complementary in expressive power \cite{Bonifati2018}. \CPQ{} is an expressive subset of {\em CQ}. 

{\em RPQ} and {\em CQ} require fundamentally different indexing and processing methods because they support fundamentally different operations: {\em RPQ} does not support conjunctions of paths and cyclic patterns; and, {\em CQ} and \CPQ{} do not support disjunctive patterns and Kleene star (i.e., transitive closure).
We focus on \CPQ{} for index design since (1) constructing {\em CQ} equivalence classes for {\em CQ}-aware indexes has impractical exponential cost \cite{Rossman08x} (even if restricted to paths), unlike \CPQ-aware indexes which have guaranteed polynomial cost, (2) every {\em CQ} can be evaluated in terms of its \CPQ{} sub-queries, and (3)
\CPQ{} covers more than 99\% of query shapes appearing in practice~\cite{bonifati2020analytical,BonifatiMT19}, even though \CPQ{} is a subset of {\em CQ}. 
An important topic beyond the scope of this paper is to study practical indexing for supporting richer languages such as the {\em Conjunctive Regular Path Queries}~\cite{Bonifati2018}.


\begin{table}[]
    \centering
    \caption{The comparison of language-aware path indexes}
    \label{tab:structuralindex}
    {\footnotesize
    \begin{tabular}{|l|l|l|}\hline
         \multicolumn{1}{|c|}{Index}& \multicolumn{1}{|c|}{Graph model} & \multicolumn{1}{|c|}{Query language}\\ \Hline
         DataGuides \cite{GoldmanW97} & Rooted semi-structured graph& RPQ  \\
         A[k]-index \cite{kaushik2002exploiting} & Rooted semi-structured  graph & RPQ  \\
         T-index \cite{MiloS99} & Rooted semi-structured  graph& RPQ  \\
         P(k)-index \cite{fletcher2009methodology}  & Tree & XPath \\
         Our index& Complex graph & CPQ  \\\hline
    \end{tabular}
    }
        \vspace{-5mm}
\end{table}

A rich literature exists on path indexing \cite{fletcher2016efficient,GouC07,Haw20111317,WongYT06,ShashaWG02,fletcher2009methodology,MiloS99,KaushikBNK02}.
As highlighted Section 1, 
there are two major approaches to path indexing: query language {\em unaware} and query language {\em aware}
(see also Section~\ref{sec:path-indexing} for formal definitions). 
Language-unaware path indexes are designed for supporting path patterns~\cite{fletcher2016efficient,ShashaWG02}. 
While effective for accelerating simple graph path queries, they are not sufficient to accelerate complex path queries arising in practice, 
as they do not leverage the richer topological structure of graphs exposed by languages such as \CPQ, as heavily used in real query workloads \cite{bonifati2020analytical}.
Examples of language-unaware indexes include a trie tree-based path index~\cite{cooper2001fast} and an RDF triple-based path index~\cite{ baolin2007hprd}.
The state-of-the-art language-unaware path index is an inverted index for label sequences~\cite{fletcher2016efficient}.

Language-aware path indexing, which does leverage richer graph structures, has been developed in the context of semi-structured, XML, and RDF data \cite{fletcher2009methodology,MiloS99,KaushikBNK02,PicalausaLFHV12,WongYT06}. 
All prior work on language-aware path indexing, however, has focused on data models and/or query languages incomparable with \CPQ{}, e.g.,
DataGuides \cite{GoldmanW97}, A[k]-index \cite{kaushik2002exploiting}, and T-index \cite{MiloS99} for {\em RPQ} on rooted semi-structured graphs, and the P(k)-index \cite{fletcher2009methodology} for XPath queries on trees.
We summarize the data models and query languages of existing language-aware path indexes and our proposal in Table \ref{tab:structuralindex}.
These language-aware path indexes are not applicable to \CPQ{} for the following three reasons.
First, existing indexes are for rooted semi-structured graphs and trees, and thus their construction methods are not applicable to arbitrary graphs.
Indeed, it is well known that reasoning about bisimulation structures on trees is much cheaper than on arbitrary graphs with cycles \cite{hellings2012efficient}.
Second, {\em RPQ} and XPath do not support cyclic query patterns and/or conjunction of paths, so their structural characterizations are inapplicable to \CPQ.
Third, all indexes except for T-index and P(k)-index are {\em vertex-based} in the sense that the indexes are built over partitions of the set of vertices in the graph. Vertex-based indexes do not support general path queries because path queries require reasoning over the start and end vertices of paths. 
For example, the index designs of A[k]-index \cite{kaushik2002exploiting} and T-index \cite{MiloS99} heavily rely on query patterns that do not include conjunctions and cycles on rooted graphs (e.g., the root must not be the targets of paths). The structures of our index are essentially different from those of A[k]-index \cite{kaushik2002exploiting} and T-index \cite{MiloS99}; ours are inverted indexes and theirs are graph-based indexes.
In summary, to the best of our knowledge no earlier indexes can be adapted for \CPQ-aware path indexing on complex graphs, and furthermore it does not follow from prior work that path-bisimulation-based indexes can be effectively constructed and used in practice.


 Methods for computing bisimulation equivalence typically focus on partitioning the {\it vertex set} of a graph \cite{aceto,luo2013}. 
 We propose here a practical method for partitioning the {\em set of paths} in a graph, which is novel in the literature.

Methods developed for exact subgraph pattern matching (i.e., {\em CQ} evaluation) can be used to process \CPQ.
Here, matching has been studied mainly under two matching semantics: isomorphic and homomorphic.
Systems for isomorphic subgraph matching, e.g., \cite{han2019efficient,HanLL13,lee2012depth,fan2020extending}, are not suitable for \CPQ{} which has homomorphic matching semantics (see Section 3.2).
Isomorphic subgraph matching methods can return incorrect results when processing \CPQ{}.
Systems for homomorphic subgraph matching including RDF engines such as RDF-3X \cite{NeumannW10}, Virtuoso \cite{erling2009rdf}, and Tentris~\cite{bigerl2020tentris}, and subgraph matching algorithms such as TurboHom++ \cite{kim2015taming} are applicable to process \CPQ. 
To the best of our knowledge, TurboHom++ and Tentris are the state-of-the-art algorithm and RDF engine for a homomorphic subgraph matching, respectively. We compare our methods with TurboHom++ and Tentris in our experimental study.


\section{Preliminaries}
\label{sec:preliminaries}

We study the evaluation of conjunctive path queries on directed edge-labeled graphs using path-based index data structures.
In this section we define these concepts. 

\subsection{Graphs, paths, and label sequences}
A {\em graph} is a triple $\graph =(\vertices, \edges, \elabels)$ where $\vertices$ is a finite set of {\em vertices} and   $\edges \subseteq \vertices \times \vertices \times \elabels$ is
a set of labeled directed {\em edges}, i.e.,  $(v,u,\elabel) \in \edges$ denotes an edge from head vertex $v$ to tail vertex $u$ with label $\elabel \in \elabels$. $\elabels$ is a finite non-empty set of labels.\footnote{For simplicity we do not consider vertex labels.  Extending our methods to accommodate labels on vertices is straightforward.}
To support traversals in the inverse direction of edges, we extend $\elabels$ with $\inverse{\elabel}$ for $\elabel\in\elabels$ and $\edges$ with $(u,v,\inverse{\elabel})$ for $(v,u,\elabel)\in\edges$.


We refer to pairs of vertices $(v, u) \in \vertices \times \vertices$ as {\em source-target vertex pairs}, where $v$ and $u$ are the sources and targets, respectively.
We define $\paths^{\leq k}$, for $k\geq 0$, to be the set of all those source-target pairs such that there is a path of length at most $k$ in \graph{} from the source of the path to its target.
We note that $\paths^{\leq k} \subseteq \vertices\times \vertices$ for any $k$. 
In the sequel, we call source-target vertex pairs as {\it s-t pairs}.

For a non-negative integer $k$, a {\em label sequence of length $k$} 
is a sequence of $k$ elements from $\elabels$ (including the inverse of labels).
We denote the set of all label sequences of length at most $k$ by  
$\elabels ^{\leq k}$ and a label sequence in $\elabels ^{\leq k}$ by $\overline{\elabel} = \langle \elabel_1, \ldots, \elabel_j \rangle$ (where $j\leq k$). 
Further, we denote by $\elabels^{\leq k}(v,u)$ the
set of all those elements $\overline{\elabel}$ of $\elabels^{\leq k}$ such that $\overline{\elabel}$ is the sequence of edge labels along a path from $v$ to $u$ in \graph. 
We define $\gamma$ as the average size of $\elabels^{\leq k}(v,u)$, over all s-t pairs $(v,u)\in\paths^{\leq k}$.

\vspace{0.1cm}
\begin{example}\label{ex:paths} 
    For $\graph_{ex}$ of Figure \ref{fig:graph}, 
    $\paths^{\leq 2}$ includes, for example, $(\mathit{ada}, \mathit{ada})$ and $(\mathit{joe},\mathit{sue})$. $\elabels^{\leq 2}(\mathit{ada},\mathit{ada})$ and $\elabels^{\leq 2}(\mathit{joe},\mathit{sue})$ include $\{ \langle \edgefont{f}, \inverse{\edgefont{f}} \rangle, \langle \edgefont{v}, \inverse{\edgefont{v}} \rangle, \langle  \inverse{\edgefont{f}},\edgefont{f} \rangle \}$ and $\{ \langle \inverse{\edgefont{f}} \rangle, \langle \edgefont{f}, \edgefont{f} \rangle,$ $\langle \edgefont{v}, \inverse{\edgefont{v}} \rangle\}$, respectively.
    \hfill{} $\square$
\end{example}



\subsection{Conjunctive path queries}
\label{ssec:cpq}
We express conjunctive path queries algebraically.  {\em Conjunctive path query} (\CPQ) expressions are all and only those built recursively from the nullary operations of identity `\identity'  and edge labels `\elabel',
using the binary operations of join `$\cmp$' and conjunction `$\cap$'.
We have the following grammar for \cpq\ expressions (for $\elabel\in\elabels$):
\begin{eqnarray*}
\cpq\!\!\!\!\! &::=&\!\!\!\!\! \identity \mid 
\elabel \mid\! \cpq \cmp \cpq \!\mid\! \cpq \cap \cpq \!\mid\! (\cpq).
\end{eqnarray*}

Let $\query\in \cpq$. 
Given graph \graph, the semantics $\eval{\query}{\graph}$
of evaluating \query\ on \graph\ is defined recursively on the structure of \query, as follows:
\begin{eqnarray*}
    \eval{\identity}{\graph} &=& \{(v, v) \mid v\in \vertices\}, \\
    \eval{\elabel}{\graph} &=& \{(v, u) \mid (v, u, \elabel)\in \edges\},\\ 
    \eval{\query_1 \cmp \query_2}{\graph} &=& 
        \{(v, u)\mid \exists m\in \vertices:
             (v, m)\in\eval{\query_1}{\graph} \\
     && ~ \qquad\qquad\qquad   \qquad \text{{\normalfont and~}} (m, u)\in\eval{\query_2}{\graph}\},\\
    \eval{\query_1 \cap \query_2}{\graph} &=& 
        \{(v, u)\mid (v, u)\in\eval{\query_1}{\graph} \text{{\normalfont ~and~}} (v,
         u)\in\eval{\query_2}{\graph}\},  \\
    \eval{(\query_1)}{\graph} &=& \eval{\query_1}{\graph}.
\end{eqnarray*}
Note that the output of a \CPQ{} is always a set of s-t pairs in $\graph$.

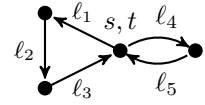
\begin{figure}[!t]
\centering
\begin{tikzpicture}[->,>=stealth',shorten >=1pt,auto,node distance=2cm,  thick,main node/.style={fill,circle,draw, minimum size=0.05cm,scale=0.5,font=\sffamily\normalsize}]

\node[main node] (2) at (0,0) [label=above:{$s,t$}] {$ $};
\node[main node] (1) at (-1,0.5) {$ $};
\node[main node] (3) at (-1,-0.5) {$ $};
\node[main node] (4) at (1.0,0) {$ $};
\path[->, every node/.style={font=\small}]
(2) edge [right] node[above] {$\elabel_1$~}   (1)
(3) edge [right] node[below] {$\elabel_3$~}   (2)
(1) edge [left]  node[left] {$\elabel_2$}   (3);
\path[<-]
(2) edge [bend right] node[below] {~~$\elabel_5$}   (4)
(4) edge [bend right]  node[above] {~~$\elabel_4$}   (2);
\end{tikzpicture}
\caption{Visual representation of the query $[(\elabel_1 \cmp \elabel_2 \cmp \elabel_3)\cap (\elabel_4 \cmp \elabel_5)] \cap id$, with output $(s, t)$.}\label{fig:graphicalnotation}
\vspace{-5mm}
\end{figure}

Figure~\ref{fig:graphicalnotation} illustrates a visual representation of a \CPQ{} query, where $s$ and $t$ denote the source and target vertices, resp.,  of paths in the query results (in this case, they are the same vertex, due to conjunction with identity).
Essentially, evaluating a \CPQ{} amounts to finding all embeddings of the pattern specified by the query into the graph.
Note that \CPQ{} has homomorphic pattern embedding semantics, as practical graph query languages such as SPARQL, G-CORE, and SQL/GQL use homomorphic semantics. 
\CPQ{} does not support conjunctions of the same label sequences.

For an expression $q\in\cpq$, we define the
\emph{diameter} $\diameter(q)$ of $q$.
Intuitively, the diameter of an expression is the maximum number edge labels to which the join
operation is applied.  
We compute query diameter as follow.
The identity operation has diameter zero; every
edge label has diameter one; 
$\diameter(q_1\cap q_2) = \max(\diameter(q_1),\diameter(q_2))$; and, $\diameter(q_1
\cmp q_2) = \diameter(q_1) + \diameter(q_2)$. 
For non-negative integer $k$, we denote by \CPQ$_k$ the set of
all expressions in \CPQ\ of diameter at most $k$.

\subsection{Language-aware and -unaware path index}\label{sec:path-indexing}
Language-aware indexing is a general methodology which leverages language-specific structural filtering in index design \cite{Bonifati2018,fletcher2009methodology,kaushik2002exploiting}.
Given a query language $L$, the basic idea of $L$-aware path indexing is to partition the set of paths in a graph \graph{} into $L$-equivalence classes and builds an index on the set of equivalence classes.
Here, an $L$-equivalence class is a set of paths in \graph{} which cannot be distinguished by any query in $L$, i.e., for every query $q\in L$, either all paths in the class appear in the evaluation of $q$ on \graph, or none of the paths appear; for evaluating $L$ the paths in the class can be processed together, instead of individually, leading to accelerated query evaluation.
\noop{
We define $L$-equivalence class as follows: 
\begin{definition}[$L$-equivalence class] \label{def:l-equivalence}
Given a query language $L$ and a graph $\graph$, an $L$-equivalence class of $\graph$ is a set of paths in $\graph$ that cannot be distinguished by any structural filtering expressed by $L$.
\end{definition}
The language-aware path indexes are a general methodology which leverages language-specific structural filtering and data models in index design \cite{Bonifati2018,kaushik2002exploiting}.
}



Language-{\em un}aware path indexes do not leverage $L$-equivalence classes.
The major drawbacks of language-unaware path indexes, relative to our \CPQ-aware path index presented in Section \ref{sec:index}, are (1) the failure to capture cyclic and conjunctive path structures, so they cannot efficiently evaluate queries with cycles and conjunctions and (2) storing the same paths multiple times in the index, leading to increased index size.
The state-of-the-art language-{\em unaware} path index~\cite{fletcher2016efficient} is an inverted index that outputs a set of paths corresponding to a given label sequence as a search key.
More precisely, 
given $\overline{\elabel} = \langle \elabel_1, \ldots, \elabel_j \rangle \in \elabels ^{\leq k}$ for some $j\leq k$, the language-unaware path index retrieves all paths associated with the label sequence.
The size of the path index~\cite{fletcher2016efficient} is $O(\gamma|\paths^{\leq k}|)$ because each path is stored $\gamma$ times on average.

\begin{figure*}[t]
	\includegraphics[width=1.0\linewidth]{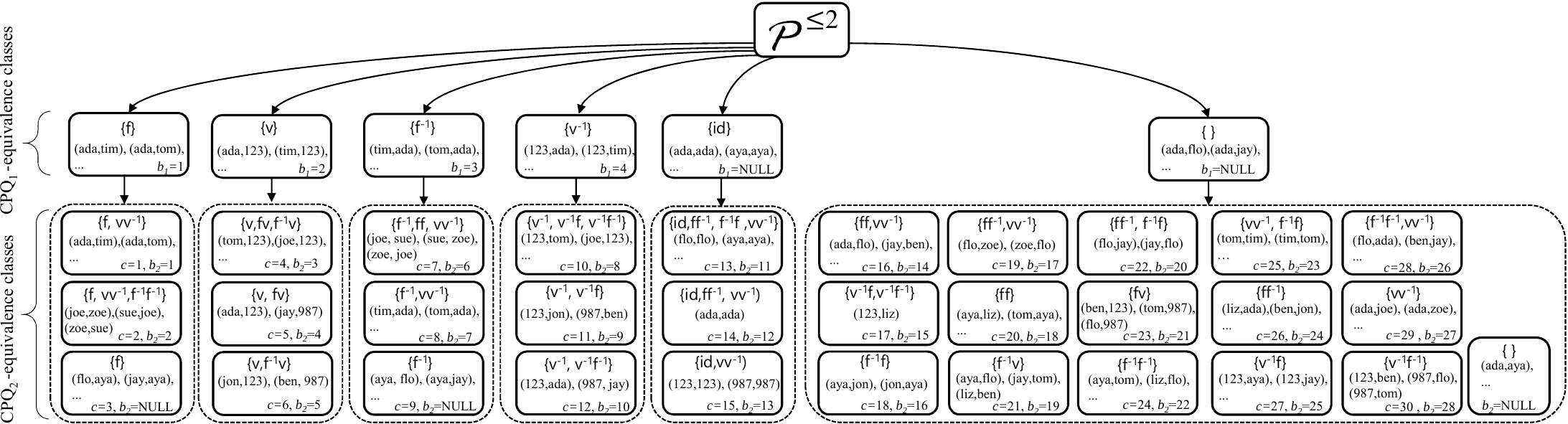}
	\vspace{-5mm}
    \caption{The \CPQ-aware index of $\graph_{ex}$ with $k=2$, where
    $c$ and $b$ indicate class and block identifiers, respectively. 
     Each equivalence class is labeled with the set of label sequences $\elabels^{\leq k}(v, u)$, for an s-t pair $(v, u)$ in the class (note that $\elabels^{\leq k}(v, u)$ is the same for all s-t pairs of the class, due to Definition 4.2).
    An arrow from block $\mathit{B}$ to block $\mathit{B}'$ (dashed rectangles, resp.) indicates that $\mathit{B}'$ (blocks within the dashed rectangles, resp.) refines $\mathit{B}$. 
    }
    \label{fig:bisimulation_example}
    \vspace{-4mm}
\end{figure*}

\section{CPQ-Aware Path Index}
\label{sec:index}

In this section, we present 
(1) our \CPQ-aware path index, \cpqindex, 
(2) an algorithm for constructing \cpqindex,
(3) an algorithm for efficient query processing with the index, 
and (4) a method for effective maintenance of the index under graph updates.

\noop{
First, we describe our overall idea of \cpqindex{} (Sec. \ref{ssec:overall}).
Second, we define \cpqindex{} (Sec. \ref{ssec:indexdef}).
Third, we explain how to construct \cpqindex{} efficiently (Sec. \ref{ssec:construction}).
Fourth, we explain an algorithm for query processing with \cpqindex{} (Sec. \ref{ssec:query}).
Finally, we explain how to update \cpqindex{} (Sec. \ref{ssec:maintenance}).
}

\subsection{Overall idea}\label{sec:idea}
The novelty of \cpqindex{} is to partition the s-t pairs $\paths^{\leq k}$ in a given graph $\graph$ into \CPQ-equivalence classes.
This is achieved by partition based on the notion {\it k-path-bisimulation}, tightly coupled to the expressive power of \CPQ$_k$ (see Theorem \ref{thm:struct}, below). 
$k$-path-bismulation satisfies the property: if two s-t pairs are $k$-path-bisimilar, the two s-t pairs are indistinguishable by any $q \in \cpq_k$.
This partition brings us two benefits: efficient pruning for CPQs and compact index representations. 


\cpqindex{} is built over \CPQ-equivalence classes in two data structures, $I_{l2c}$ and $I_{c2p}$ (Sec.~\ref{ssec:indexdef}),
which are practically constructed (Sec. \ref{ssec:construction}) and maintainable while guaranteeing the correctness of query results (Sec.~\ref{ssec:maintenance}).  
Intuitively, $I_{l2c}$ is a map from label sequences to \CPQ-equivalence classes and $I_{c2p}$  is a map from \CPQ-equivalence classes to s-t pairs. 
The index is used to evaluate queries in two stages (Sec.~\ref{ssec:query}).
In the first stage, the query is processed over the set of \CPQ-equivalence classes. 
This stage allows us to filter out paths which will not contribute to the query result. 
In the second stage, standard query processing proceeds on the s-t pairs contained in the \CPQ-equivalence classes.
The main advantage of our new query processing algorithm is that we compare \CPQ-equivalence classes while computing conjunctions of path queries instead of comparing paths themselves, resulting in significantly accelerated \CPQ{} evaluation.

\subsection{Index definition}
\label{ssec:indexdef}

\cpqindex{} is based on \CPQ-equivalence class under the notion of {\it $k$-path-bisimulation}.
We choose this notion as it captures precisely the expressive power of \CPQ$_k$.
Intuitively, s-t pairs $(v, u)$ and $(x, y)$ are $k$-path-bisimilar when all steps along any paths in the graph of length at most $k$ from $v$ to $u$ and from $x$ to $y$ can be performed in unison, every move along the way in one of the paths being mimicable in the other.


\begin{definition}[$k$-path-bisimulation] \label{def:bisim}
Let \graph\ be a graph, $k$ be a non-negative integer, and
$v, u, x, y\in\vertices$.
The s-t pairs $(v, u)$ and $(x, y)$ are {\em $k$-path-bisimilar}, denoted
$(v, u) \kbisimilar (x, y)$, if and only if 
\begin{enumerate}
    \item $v=u$ if and only if $x=y$; (i.e., cycle or not)
        \noop{
    \begin{enumerate}
        \item $v=u$ if and only if $x=y$; 
    \end{enumerate} 
}
    \item if $k > 0$, then for each $\elabel\in \elabels$,
    \begin{enumerate}
        \item if $(v, u, \elabel)\in\edges$, then $(x, y, \elabel)\in\edges$; 
            and, if $(u, v, \elabel)\in\edges$, then $(y, x, \elabel)\in\edges$;
        \item if $(x, y, \elabel)\in\edges$, then $(v, u, \elabel)\in\edges$; 
            and, if $(y, x, \elabel)\in\edges$, then $(u, v, \elabel)\in\edges$; and,
    \end{enumerate} 
    \item if $k> 1$, then 
    \begin{enumerate}
        \item for each $m\in \vertices$, if $(v, m)$ and $(m, u)$
            are in $\paths^{\leq k-1}$, then there exists $m'\in
            \vertices$ such that $(x, m')$ and $(m', y)$ are in $\paths^{\leq k-1}$, and, furthermore,  
            $(v, m) \ibisimilar{k-1} (x, m')$ and
            $(m, u)$ $\ibisimilar{k-1} (m', y)$;
        \item for each $m\in \vertices$, if $(x, m)$ and $(m, y)$
            are in $\paths^{\leq k-1}$, then there exists $m'\in
            \vertices$ such that $(v, m')$ and $(m', u)$ are in $\paths^{\leq k-1}$,  and, furthermore, 
            $(x, m) \ibisimilar{k-1} (v, m')$ and
            $(m, y)$ $\ibisimilar{k-1} (m', u)$. 
    \end{enumerate} 
\end{enumerate}
\end{definition} 


$k$-path-bisimulation is a structural characterization of the expressive power of \CPQ$_k$, in the following sense \cite{fletcher2015similarity}.
\begin{theorem} \label{thm:struct}
Let \graph\ be a graph, $k$ be a non-negative integer, and
$v, u, x, y\in\vertices$.  
If $(v, u) \kbisimilar (x, y)$, then for every $\query\in \cpq_k$ it holds that
$(v, u)\in \eval{\query}{\graph}$ if and only if
$(x, y)\in \eval{\query}{\graph}$. 
\end{theorem}

Towards leveraging Theorem \ref{thm:struct} for \cpqindex{} design, 
we define the notion of a \CPQ$_k$-equivalence class based on the $k$-path-bisimulation.
Partitioning $\paths^{\leq k}$ into \CPQ$_i$-equivalence classes 
provides a basic building block of our index. 

\begin{definition}
Let $\graph$ be a graph, $v,u \in \vertices$, $i$ be a non-negative integer, and 
    $(v,u) \in \paths^{\leq i}$. The \CPQ$_i$-equivalence class of $(v, u)$ is the set 
\begin{multline*}
    [(v, u)]_i(\graph) = \{(x,y) \mid x,y \in \vertices 
    \mathrm{~and~} (v,u) \approx_i (x,y) \}.
\end{multline*}
We denote equivalence class blocks by $B$ and a set of blocks by 
$\blocks$. We define $\blocks_i(\graph) =  \{ [(v,u)]_i(\graph) \mid v,u \in \vertices \}.$
\end{definition}

As a corollary of Theorem \ref{thm:struct}, we have that query processing is tightly coupled to $\blocks_k(\graph)$. 
\begin{corollary} \label{cor:block}
Let \graph\ be a graph, $k$ be a non-negative integer, and
$\query\in \cpq_k$.  There exists $\blocks\subseteq \blocks_k(\graph)$ such that
$\eval{q}{\graph} = \bigcup_{\mathit{B}\in \blocks} \mathit{B}.$
\end{corollary}

s-t pairs are disjointly partitioned into blocks in $\blocks_k$ based on the notion of $k$-path-bisimulation.
Towards leveraging Corollary \ref{cor:block} for query processing, we assign a {\em class identifier} $c$ to each block in $\blocks_k(\graph)$, and we define $\classes$ as the set of class identifiers.
We also define $\paths (c)\subseteq \paths^{\leq k}$ and $\classes(\overline{\elabel}) \subseteq \classes$ as the set of s-t pairs that belong to $c$, and the set of class identifiers that belong to $\overline{\elabel}$ (i.e., those equivalence classes whose elements are in the evaluation result of $\overline{\elabel}$ interpreted as a query), respectively.
We can now define \cpqindex{} based on the \CPQ$_k$-equivalence class of $\graph$.

\begin{definition}[CPQ-aware Index \cpqindex{}]
    Given $\blocks_k(\graph)$, \cpqindex{} $I_k$ is a pair of data structures $I_{l2c}$ and $I_{c2p}$ such that 
$I_{l2c}$  maps
    label sequences in $\elabels^{\leq k}$ to sets of class identifiers and
$I_{c2p}$ maps
    class identifiers to sets of s-t pairs in $\paths^{\leq k}$, as follows:
\begin{eqnarray*}
    I_{l2c}(\overline{\elabel}) &=& \{ c \mid c \in \classes (\overline{\elabel}) \},  \\
    I_{c2p}(c)                  &=& \{ (v,u) \mid (v,u) \in \paths(c)  \}.
\end{eqnarray*}
\end{definition}

\cpqindex{} is essentially an inverted index to find the set of s-t pairs associated with given label sequences through class identifiers. 
We note that s-t pairs are not stored  in \cpqindex{} if they are not connected by path with at most $k$.
It enables us to efficiently find the set of s-t pairs that satisfy \CPQ s thanks to the \CPQ$_k$-equivalence classes.

\begin{theorem}
    The size of \cpqindex{} is $O(\gamma | \classes |+|\paths^{\leq k}|)$.
\end{theorem}
 {\it Proof:} 
 $I_{l2h}$ stores the set of class identifiers associated with each label sequence. 
Each class identifier appears on average $\gamma$ times in $I_{l2h}$. 
Thus, the size of $I_{l2h}$ is $O(\gamma | \classes |)$.
In $I_{h2p}$, since each path is stored as single entry, the size of  $I_{h2p}$ is $O(|\paths^{\leq k}|)$.
Therefore, the size of \cpqindex{} is $O(\gamma | \classes |+|\paths^{\leq k}|)$.
\hfill{} $\square$

The size of \cpqindex{} $O(\gamma | \classes |+|\paths^{\leq k}|)$ is not larger than that of language-unaware path index~\cite{fletcher2016efficient} $O(\gamma|\paths^{\leq k}|)$ because $|\classes |$ is at most $|\paths^{\leq k}|$. 
Each s-t pair in \cpqindex{} is associated with a single class, while each s-t pair in state-of-the-art path indexes can be associated with multiple label sequences, as we observed in Section~\ref{sec:path-indexing}.
\cpqindex{} achieves smaller size than the state-of-the-art language-unaware path index.

\vspace{0.1cm}
\begin{example}\label{ex:bisimulation}
Figure~\ref{fig:bisimulation_example} shows \CPQ$_k$-equivalent classes of $\graph_{ex}$ of Figure~\ref{fig:graph} for $k=2$.
The first and second rows of Figure~\ref{fig:bisimulation_example} depict $\blocks_1(\graph_{ex})$ and $\blocks_2(\graph_{ex})$, respectively.

Suppose that we find s-t pairs of paths labeled with both $\edgefont{\inverse{f}}$ and $\edgefont{ff}$.
In \cpqindex, $I_{l2c}(\edgefont{\inverse{f}})$ and $I_{l2c}(\edgefont{ff})$ return $\{7, 8, 9\}$ and $\{7, 16, 20\}$, respectively.
From these results, we can see that the s-t pairs of paths labeled with $\{ \edgefont{\inverse{f}, ff}\}$ belongs only to $7$. We do not need to check s-t pairs in other blocks.
\hfill{} $\square$


\end{example}


\subsection{Index construction}
\label{ssec:construction}

We describe an efficient construction method for \cpqindex{} defined in Section~\ref{ssec:indexdef}.
The main contribution here is that we develop an efficient algorithm to compute \CPQ$_k$-equivalence class whose time and space complexity are polynomial for our index design.
After computing the \CPQ$_k$-equivalence class, we construct $I_{k} = (I_{l2c}, I_{c2p})$.

We use a bottom-up approach for computing \CPQ$_k$-equivalence class.
We here note that \CPQ$_k$-equivalence class in our index construction is not exactly same as the original definition. Our index does not need to distinguish paths with conjunctions divided at different locations if two paths are merged at targets.
The bottom-up approach is suitable for computing \CPQ$_k$-equivalence class in our index design.

A straightforward algorithm is that (1) for computing $1$-path-bisimulation, it checks all s-t pairs if they are connected edges with the same labels and cyclic patterns (i.e., self-loop), (2) it obtains pairs of paths that are connected 2 length path and computes 2-path-bisimulation, and (3) it repeats similar process of the second one until obtaining $k$-path-bisimulation.
However, this is inefficient because we need to compute $i$-path bisimulation for all s-t pairs.

The idea of our algorithm leverages the characteristics of bisimilar paths, which (1) paths uniquely belong to blocks in $\blocks_i$ and (2) $i$-path-bisimilation refines $(i-1)$-path-bisimilation.
We assign a {\it block identifier} $b_i(v,u)$ to each block $[(v,u)]_i$ and $(v,u)$ of $\blocks_i$.
Each s-t pair $(v, u) \in \paths^{\leq k}$ has an associated sequence of $k$ block identifiers $\langle b_1(v, u), \ldots, b_k(v,u) \rangle$.
It is easy to establish that $k$-path-bisimilar s-t pairs are uniquely identified by their common sequences.

Based on the above idea, our algorithm effectively skips both computing $i$-path-bisimulation and assigning block identifiers to the s-t pairs that have no paths.
For obtaining block identifiers of $\blocks_i$, we join two s-t pairs in $\blocks_{i-1}$ and $\blocks_1$.
We can assign the same block identifiers to s-t pairs in $\blocks_{i}$ if (1) joined s-t pairs have the same block identifiers in $\blocks_{i-1}$ and $\blocks_1$ and (2) both s-t pairs are cycle or not.
Furthermore, we can skip assigning block identifiers to blocks in $\blocks_i$ if the s-t pairs in $\blocks_i$ are not connected at $i$ length paths because $k$-path-bisimilar s-t pairs are uniquely identified even if $b_i(v,u) = \mathit{Null}$.

After computing the \CPQ$_k$-equivalence class, \cpqindex{} is constructed in a simple way.
It generates class identifiers from sequences of block identifiers, and then insert a pair of $\overline{\elabel}$ and $c \in \classes (\overline{\elabel})$ into $I_{l2c}$ and a pair of class identifier $c$ and $(v, u) \in \paths(c)$ into $I_{c2p}$.
Note that the set of source-target paths with the same class identifier has the same label sequence due to the definition of $k$-path-bisimulation.


Algorithms \ref{alg:bis} and \ref{alg:construct} show pseudo-code for computing the \CPQ$_k$-equivalence class and constructing \cpqindex, respectively. 
In Algorithm \ref{alg:bis}, we sort elements of $\mathbb{S}^i$ so that $i$-path-bisimilar s-t pairs are sequentially listed for efficiently assigning the block identifiers. 
In Algorithm \ref{alg:construct}, we generate class identifier $c$ of $(v,u)$ by using a hash function for each class $\langle b_1(v, u), \ldots, b_k(v,u) \rangle$. If two s-t pairs have the same class, it assigns the same class identifiers to the two s-t pairs.

\begin{algorithm}[!t] 
	\caption{Computing \CPQ$_k$-equivalence class}	\label{alg:bis}
	{\small
		\DontPrintSemicolon
			    \SetKwInOut{Input}{input}
	            \SetKwInOut{Output}{output}
	            \Input{Graph $\mathcal{G}$, natural number $k$}
	            \Output{Set of blocks $[\graph]_k$}
            	{\bf procedure} {\sc cpqPathPartition}($\graph$, $k$)\\
            	$\mathbb{S}^i_{(v,u)} = \emptyset$ for $i=1,\ldots,k$ and $\forall (v,u) \in (\paths^{\leq i} - \paths^{\leq i-1}) \cup (\paths^{\leq i} \cap \paths^{\leq i-1})$;\\
            	\For{$e=(v,u,\elabel) \in \edges$}{
            	    $\mathbb{S}^1_{(v,u)} \leftarrow \mathbb{S}^1_{(v,u)} \cup \{\elabel\}$;\\
            	}
            	Sort $\mathbb{S}^1$ according to edge labels and $(v,u)$;\\
            	Set $b_1(v,u)$ for $\forall (v,u) \in \paths^{\leq 1}$ as $\blocks_1$;\\
            	\For{$i = 2, \ldots, k$}{
            	    \For{$\forall \mathbb{S}^{i-1}_{(v,m)}$}{
            	        \For{$\forall \mathbb{S}^{1}_{(m,u)}$}{
            	            $\mathbb{S}^{i}_{(v,u)} \leftarrow \mathbb{S}^{i}_{(v,u)} \cup$ \{$b_{i-1}(v,m)$, $b_1(m,u)$\};\\
            	        }
            	    }
            	    Sort $\mathbb{S}^i$ according to block identifiers and $(v,u)$;\\
            	   	Set $b_i(v,u)$ for $\forall (v,u) \in (\paths^{\leq i} - \paths^{\leq i-1})\cup(\paths^{\leq i} \cap \paths^{\leq i-1})$ as $\blocks_i$;\\
            	   	{\bf if} $i\neq2$ {\bf then} Clear $\mathbb{S}^{i-1}$:\\
              }
              {\bf return} $[\graph]_k =\{\blocks_1,\ldots,\blocks_k\}$;\\
              {\bf end procedure}
    }
\end{algorithm}

\begin{algorithm}[!t] 
	\caption{Construction of \cpqindex}	\label{alg:construct}
	{\small
		\DontPrintSemicolon
			    \SetKwInOut{Input}{input}
	            \SetKwInOut{Output}{output}
	             \SetKwFunction{BFS}{BFS}
	            \Input{Graph $\mathcal{G}$, natural number $k$}
	            \Output{\cpqindex{} $I_{k}=\{ I_{c2p}, I_{l2c}\}$}
            	{\bf procedure} {\sc Construction}($\graph$, $k$)\\	
            	$[\graph]_k$ $\leftarrow$ {\sc cpqPathPartition}($\graph$, $k$);\\
            	\For{$(v,u) \in \paths^{\leq k}$}{
            	    $c \leftarrow hash(\langle b^1_{v,u}, \ldots, b^{k}_{v,u}\rangle)$;\\
            	   \If{$c$ is $\mathit{Null}$}{
            	        $c \leftarrow c_{new}$; \\     
            	        $hash(\langle b^1_{v,u}, \ldots, b^{k}_{v,u}\rangle) \leftarrow c$ ;\\
            	        $\classes \leftarrow \classes \cup \{c\};$\\
            	        Update $c_{new}$;\\
            	   }
            	   $I_{h2p}.append(c,(v,u))$;\\
            	}
            	\For{$c \in \classes$}{
            	    \For{$(v,u) \in I_{c2p}(c)$}{
                	    \For{$\overline{\elabel} \in \elabels^{\leq k}(v,u)$}{
                	        $I_{l2c}.append(\overline{\elabel}, c)$;\\
                	    }
            	    }
            	}
            	sort $(v,u)$ in $I_{c2p}$ and $c$ in $I_{l2c}$;\\
              {\bf return} $I_{k}$;\\
              {\bf end procedure}
    }
\end{algorithm}

\noop{
\subsubsection*{Algorithm}
Algorithms \ref{alg:construct} and \ref{alg:bis} show pseudo-codes for constructing \cpqindex{} and computing $k$-path-bisimulation, respectively.
We first compute $k$-path-bisimulation partition (line 2 Algorithm \ref{alg:construct}).
Algorithm \ref{alg:bis} computes $k$-path-bisimulation by repeatedly computing blocks $B^0, \cdots B^k$.
After computing $k$-path bisimulation partition, we assign class identifiers (lines 5--8  Algorithm \ref{alg:construct}). 
Then, we insert the set of class identifiers and paths into $I_{c2p}$ and $I_{l2c}$, respectively (lines 9 and 13 Algorithm \ref{alg:construct}).
Finally, it sorts the set of paths and class identifiers, respectively (line 14 Algorithm \ref{alg:construct}).
}

We here describe the time and space complexity for constructing \cpqindex.
$d$ indicates the maximum vertex degree.

\begin{theorem}[Time complexity]\label{thm:timecomplexity}
Given a graph $G$ and positive number $k$, the time complexity of index construction is $O( k(d|\paths^{\leq k}| + |\paths^{\leq k}| \log |\paths^{\leq k}|)+\gamma|\classes| \log \gamma|\classes|)$.
\end{theorem}
{\it Proof:} The algorithm for constructing \cpqindex{} has two steps (1) computing $[\graph]_k$ and (2) constructing $I_k = (I_{l2c}, I_{c2p})$. 
    For computing ${[\graph]}_k$, the algorithm enumerates the set of block identifiers for each path, which takes $O(d|\paths^{\leq k}|)$. Then, it compares the block identifiers by a sorting algorithm, which takes $O(|\paths^{\leq k}| \log |\paths^{\leq k}|)$. Since it repeats $k$ times, it takes $O(k(d|\paths^{\leq k}| + |\paths^{\leq k}| \log |\paths^{\leq k}|))$.
    For constructing $I_{_k} = (I_{l2c}, I_{c2p})$,
    it sorts the set of pairs of label sequences and class identifiers for $I_{l2c}$ and the set of pairs of class identifiers and paths for $I_{c2p}$.
    Since these sizes are $O(|\paths^{\leq k}|)$ and $O(\gamma|\classes|)$, resp., it takes $O(|\paths^{\leq k}|\log |\paths^{\leq k}|)$ and $O( \gamma|\classes| \log \gamma|\classes|)$, resp.
    Thus, the total time complexity is  $O(k(d|\paths^{\leq k}| + |\paths^{\leq k}| \log |\paths^{\leq k}|)+\gamma|\classes| \log \gamma|\classes|)$.
\hfill{} $\square$

\begin{theorem}[Space complexity]\label{thm:spacecomplexity}
    Given a graph $\graph$ and positive number $k$, the space complexity of index construction is $O((k+d)|\paths^{\leq k}|+\gamma|\classes|)$.
\end{theorem}
{\it Proof: } The algorithm stores block identifiers (or label sequences) for paths in $(\paths^{\leq i} - \paths^{\leq i-1})\cup (\paths^{\leq i} \cap \paths^{\leq i-1})$ and the number of block identifiers for each path is at most $d$. 
Thus, the size of $\mathbb{S}^i$ is $O(d|\paths^{\leq k}|)$.
Additionally, it stores $k$ sets of block identifiers (i.e., $\blocks_1,\ldots, \blocks_k$). Since the size of each set is $O(|\paths^{\leq k}|)$, the size is totally $O(k|\paths^{\leq k}|)$. 
To store the index, it takes $O(\gamma|\classes|+|\paths^{\leq k}|)$.
Therefore, its space complexity is $O((k+d)|\paths^{\leq k}|+\gamma|\classes|)$.\hfill{} $\square$

\vspace{0.1cm}
\begin{example}\label{ex:bisimulation}
We explain how to construct \cpqindex{} by using Figure~\ref{fig:bisimulation_example}.
Our algorithm first computes the 1-equivalence classes according to edge labels.
It assigns block identifiers to blocks with $\{\edgefont{f}\}$,  $\{\edgefont{v}\}$, $\{\edgefont{\inverse{f}}\}$, and $\{\edgefont{\inverse{v}}\}$, while it does not assign block identifiers to blocks with $\{\edgefont{id}\}$ and  $\{ \}$ because the s-t pairs are not connected one length paths (i.e., edges).

Then, it joins two edges to obtain paths with two length, for instance, it joins  $(\mathit{ada}, \mathit{123})$ and $(\mathit{123}, \mathit{tim})$, and then obtains $(\mathit{ada}, \mathit{tim})$ with $\edgefont{v\inverse{v}}$.
Similarly, it joins $(\mathit{ada}, \mathit{123})$ and $(\mathit{123}, \mathit{tom})$, and then obtains $(\mathit{ada}, \mathit{tom})$. 
Since joined paths of $(\mathit{ada}, \mathit{tim})$ and $(\mathit{ada}, \mathit{tom})$ are from the same s-t pairs of blocks (i.e., $(\mathit{123}, \mathit{tim})$ and $(\mathit{123}, \mathit{tom})$ belong to the same block), $(\mathit{ada}, \mathit{tim})$ and $(\mathit{ada}, \mathit{tom})$ belong to the same block.
We do not assign block identifiers to blocks with $\{\edgefont{f}\}$, $\{\edgefont{\inverse{f}}\}$, and $\{ \}$ because the s-t pairs in the block are not connected two length path.

s-t pairs in the same equivalent classes have the same sequences of block identifiers. 
Even if we do not assign block identifiers to some blocks, we can identify \CPQ$_2$-equivalence classes.
For instance, $\{ \edgefont{f}\}$ in CPQ$_2$-equivalence class has sequence $\langle 1, \mathit{Null} \rangle$, and other blocks do not have the same sequence.

Our construction algorithm effectively reduces the index construction costs because most s-t pairs have no paths.
In such small example of Figure~\ref{fig:bisimulation_example}, we can skip computing many bisimilar s-t pairs. In this example, the possible number of s-t pairs is 196 and the number of s-t pairs that are connected paths at most two length is 150. 
As graph sizes increase, the effectiveness of our index construction algorithm increases.
\hfill{} $\square$
\end{example}

\subsection{Query processing with \cpqindex{}}
\label{ssec:query}

We accelerate query processing by using \cpqindex, instead of the original graph. 
The effective use of classes mitigates the cost of unnecessarily comparing paths which do not participate in the query result.

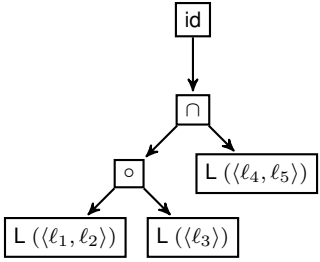
\begin{figure}[t]
  \centering
\begin{tikzpicture}[->,>=stealth',shorten >=1pt,auto,node distance=1.2cm,  thick,main node/.style={rectangle,draw,minimum size=0.1cm, font=\sffamily\footnotesize}]
  \node[main node] (0)   {{\sf id}};
  \node[main node] (1) [below of =0]  {$\cap$};
    \node[main node] (2) [below right of=1] {{\sf L} $(\langle \elabel_4 , \elabel_5\rangle)$};
  \node[main node] (3) [below left of=1]  {$\cmp$};
    \node[main node] (4) [below right of=3] {{\sf L} $(\langle \elabel_3\rangle)$};
    \node[main node] (5) [below left of=3]  {{\sf L} $(\langle \elabel_1, \elabel_2\rangle)$};
  \path[->]
    (0) edge [right] node[above] {$ $}   (1)
    (1) edge [right] node[above] {$ $}   (2)
    (1) edge [left]  node[below] {$ $}   (3)
    (3) edge [left]  node[below] {$ $}   (4)
    (3) edge [left]  node[below] {$ $}   (5);
\end{tikzpicture}
    \caption{Parse tree of query $[(\elabel_1 \cmp \elabel_2 \cmp \elabel_3)\cap (\elabel_4 \cmp \elabel_5)] \cap id$, when $k=2$.
    Here, $\elabel_1 \cmp \elabel_2 \cmp \elabel_3$ is processed left to right, with an index look up $\langle \elabel_1, \elabel_2\rangle$ joined with a look up of $\langle \elabel_3\rangle$.}
\label{fig:querytree}
    \vspace{-4mm}
\end{figure}

Our query processing method builds a parse tree according to a given query $\query\in\cpq$ and \cpqindex{} (see Figure \ref{fig:querytree} for an example) and evaluates the query following the parse tree.
Each node of the parse tree represents a logical operation of \CPQ: {\sc LookUp} (i.e., given label sequence $\overline{\elabel} \in \elabels ^{\leq k}$, find the corresponding set of class identifiers by $I_{l2c}$), {\sc Conjunction}, {\sc Join}, and {\sc Identity}.
Since {\sc LookUp} nodes depend on the length $k$ of \cpqindex, we split label sequences longer than $k$ into sub-label sequences whose sizes are at most $k$.
This method derives an execution plan, with index {\sc LookUp} as leaf nodes. 
We process starting from the root node of \query, recurring on the left and right, as necessary.
Further query optimization is an interesting rich topic for future research.

Our query processing method, in particular, accelerates {\sc Conjunction} and {\sc Identity}.
In the case of {\sc Conjunction}, we can efficiently compute the conjunction of the two sub queries without directly comparing the corresponding set of s-t pairs in the graph.
We compare class identifiers obtained by two sub queries to obtain a set of s-t pairs that satisfy both of the sub queries.
The correctness of conjunction evaluation is based on the following proposition:

\begin{proposition}[\sc{Conjunction} correctness]
Given two sets of class identifiers $\sclasses$ and $\sclasses'$, the set of s-t pairs
$\{(v, u)\mid \forall c\in \sclasses, c'\in \sclasses': (v, u)\in\paths(c)  \text{{\normalfont ~and~}} (v, u)\in\paths(c') \}$ is the same as $\{(v, u)\mid \forall c\in \sclasses \cap \sclasses': (v, u)\in\paths(c)\}$.
\end{proposition}
 
In the case of {\sc Identity}, since $k$-path-bisimilar s-t pairs are partitioned according to their cyclic patterns, we can evaluate  {\sc Identity} by just checking the first s-t pairs in the set of s-t pairs of class identifiers.
These processes decrease computation cost significantly because the number of class identifiers $|\classes|$ is much smaller than that of paths $|\paths^{\leq k}|$ (see Table \ref{tab:numhistory} in experimental study).


Additionally, we use optimization techniques to reduce computation cost while guaranteeing correctness.
First, we use a sorted merge join as a physical operator for {\sc Conjunction} and {\sc Join}, as \cpqindex{} stores sorted class identifiers and s-t pairs. 
Second, in {\sc Identity} since we can optimize $\query \cmp \identity = \query$, we handle only $\query \cap \identity$ as {\sc Identity}. 
Third, {\sc Identity} is executed with the other three operators to avoid inserting the s-t pairs that are deleted by {\sc Identity}.

We describe the time complexity of query processing by using \cpqindex.
\begin{theorem}[Time complexity]\label{thm:timecomplexity_query}
Let assume a given query $q$ consisting of $\alpha_1$ times of {\sc Join} and $\alpha_2$ times of {\sc Conjunction}. 
Given graph $\graph$ and index $I_k$, the time complexity of processing $q$ is $O((\alpha_1+\alpha_2) (\max ((d^k)^{\alpha_1}|\spaths_q|, |V|^2) \log (\max((d^k)^{\alpha_1}|\spaths_q|, |V|^2)))$ if $\alpha_1 > 0$ or $O(\alpha_2 |\sclasses_q|)$ if $\alpha_1 = 0$, where the maximum numbers of paths $|\spaths_q|$ and class identifiers $|\sclasses_q|$ returned by a single {\sc LookUp}, and the maximum degree $d$.
\end{theorem}
{\it Proof:} The query processing consists of two main process;  {\sc Join} and {\sc Conjunction}. Other operations are ignorable due to small costs.
First, in {\sc Join}, it takes  $O(|\paths_q| \log |\paths_q|)$ for the first {\sc Join}. The number of paths could increase at most ${d^k}|\paths_q|$ after the first  {\sc Join}. When $q$ consists of $\alpha_1$ times of {\sc Join}, the cost of  {\sc Join} is $O(\alpha_1 (\max ((d^k)^{\alpha_1}|\paths_q|, |V|^2) \log \max ((d^k)^{\alpha_1}|\paths_q|, |V|^2))$.
Second, in {\sc Conjunction} before {\sc Join}, it takes $O(|\sclasses_q|)$ because it compares the number overlapped class identifiers. The number of class identifiers after {\sc Conjunction} does not increase, so the costs of {\sc Conjunction}  $O(\alpha_2|\sclasses_q|)$.
If {\sc Conjunction} follows {\sc Join}, our algorithm sorts paths and check overlapped paths. It takes $O(\alpha_2 (\max (d^k)^{\alpha_1}|\paths_q|, |V|^2) \log (\max (d^k)^{\alpha_1}|\paths_q|, |V|^2))$. 

Therefore, the time complexity of processing $q$ is $O((\alpha_1+\alpha_2) (\max ((d^k)^{\alpha_1}|\spaths_q|, |V|^2) \log (\max((d^k)^{\alpha_1}|\spaths_q|, |V|^2)))$ if $\alpha_1 > 0$ or $O(\alpha_2 |\sclasses_q|)$ if $\alpha_1 = 0$.
\hfill{} $\square$

Algorithms~\ref{alg:querywhole} and \ref{alg:queryoperation} show pseudocode for our query processing method and operations, respectively.
$\spaths$ and $\sclasses$ denote the sets of s-t pairs and class identifiers that are found during query processing, respectively.
Our query processing algorithm repeatedly traverses the nodes on a given query tree. 
\begin{algorithm}[!t] 
	\caption{Query processing}	\label{alg:querywhole}
	{\small
		\DontPrintSemicolon
			    \SetKwInOut{Input}{input}
	            \SetKwInOut{Output}{output}
	             \SetKwFunction{Expand}{Expand}
	            \Input{Node on query tree $q$, structural index $I_{k}$}
	            \Output{set of paths $\spaths$, set of classes $\sclasses$}
            	{\bf procedure}  {\sc Evaluation}($q$, $I_{k}$)\\	
        	    \If{operation of $q$ is  {\sc LookUp}}{
        	        {\bf return} {\sc LookUp ($q.\overline{\elabel}$, $I_{k}$);}
        	    }
        	    \ElseIf{operation of $q$ is {\sc LookUP} with {\sc Identity}}{
            	    {\bf return} {\sc LookUpId ($q.\overline{\elabel}$, $I_{k}$);}
        	    }
            	\Else{
            	    $\spaths _l , \sclasses _l \leftarrow$ {\sc Evaluation}($q_l$, $I_{k}$);\\
                    $\spaths _r, \sclasses_r \leftarrow$ {\sc Evaluation}($q_r$, $I_{k}$);\\

                    \If{operation of $q$ is {\sc Join}}{
                	    {\bf return} {\sc Join} ($\spaths _l, \spaths _r, \sclasses_l, \sclasses_r$, $I_{k}$);
                	}
            	    \ElseIf{operation of $q$ is {\sc Conjunction}}{
            	        {\bf return} {\sc Conjunction} ($\spaths _l, \spaths _r, \sclasses_l, \sclasses_r$, $I_{k}$);
            	    }
            	    \ElseIf{operation of $q$ is {\sc Join} with {\sc Identity}}{
                	    {\bf return} {\sc JoinId} ($\spaths _l, \spaths _r, \sclasses_l, \sclasses_r$, $I_{k}$);
                	}
            	    \ElseIf{operation of $q$ is {\sc Conjunction} with {\sc Identity}}{
            	        {\bf return} {\sc ConjunctionId} ($\spaths _l, \spaths _r, \sclasses_l, \sclasses_r$, $I_{k}$);
            	    }
              }
              \If{$q$ is the root of query tree}{
                $\spaths \leftarrow \spaths \cup I_{c2p}(c)$ for all $c \in \sclasses$;
              }
              {\bf return} $\spaths, \sclasses$;\\
              {\bf end procedure}
    }
\end{algorithm}

\begin{algorithm}[!h] 
	\caption{Operations}	\label{alg:queryoperation}
		\DontPrintSemicolon
		{\small
              {\bf procedure} {\sc LookUp($\overline{\elabel}$, $I_{k}$)}\\
              {\bf return} $\emptyset, I_{c2p}(\overline{\elabel})$;\\
              \hrulefill \\
              {\bf procedure} {\sc Join}($\spaths _l, \spaths _r, \sclasses _l, \sclasses _r$, $I_{k}$)\\
              $\spaths_l \leftarrow \{(v,u) \mid (v,u) \in I_{c2p}(c) \land c \in \sclasses _l \}$;\\
              $\spaths_r \leftarrow \{(v,u) \mid (v,u) \in I_{c2p}(c) \land c \in \sclasses _r \}$;\\
              $\spaths \leftarrow \{(v,y)\mid (v,u) \in \spaths _l \land (x,y) \in \spaths _r \land u=x \}$\\
              {\bf return} $\spaths, \emptyset$;\\
              \hrulefill \\
              {\bf procedure} {\sc Conjunction}($\spaths _l, \spaths _r, \sclasses _l, \sclasses _r$, $I_{k}$)\\
              \If{$\sclasses _l \neq \emptyset$ and $\sclasses _r \neq \emptyset$}{
                $\sclasses \leftarrow \sclasses_l \cap \sclasses_r$;\\
                {\bf return} $\emptyset, \sclasses$;\\
              }
              \Else{
                \If{$\spaths _l \neq \emptyset$} {
                    $\spaths _l \leftarrow \{(v,u) \mid (v,u) \in I_{c2p}(c) \land c \in \sclasses _l \}$;\\
                }
                \If{$\spaths _r \neq \emptyset$}{ 
                    $\spaths _r \leftarrow \{(v,u) \mid (v,u) \in I_{c2p}(c) \land c \in \sclasses _r \}$;\\
                }
                $\spaths \leftarrow \spaths _l \cap \spaths _r$;\\
                {\bf return} $\spaths, \emptyset$;
              }
              \hrulefill \\
              {\bf procedure} {\sc JoinId}($\spaths _l, \spaths _r, \sclasses _l, \sclasses _r$, $I_{k}$)\\
              $\spaths_l \leftarrow \{(v,u) \mid (v,u) \in I_{c2p}(c) \land c \in \sclasses _l \}$;\\
              $\spaths_r \leftarrow \{(v,u) \mid (v,u) \in I_{c2p}(c) \land c \in \sclasses _r \}$;\\
              $\spaths \leftarrow \{(v,y)\mid (v,u) \in \spaths _l \land (x,y) \in \spaths _r \land u=x \land v=y\}$;\\
              {\bf return} $\spaths, \emptyset$;\\
              \hrulefill \\
              {\bf procedure} {\sc ConjunctionId}($\spaths _l, \spaths _r, \sclasses _l, \sclasses _r$, $I_{k}$)\\
              \If{$\sclasses _l \neq \emptyset$ and $\sclasses _r \neq \emptyset$}{
                $\sclasses \leftarrow \{c \mid c \in \sclasses_l \land c \in \sclasses_r \land (v,u) \in I_{c2p}(c) \land v=u\}$;\\
                {\bf return} $\emptyset, \sclasses$;\\
              }
              \Else{
                \If{$\spaths _l \neq \emptyset$} {
                    $\spaths _l \leftarrow \{(v,u) \mid (v,u) \in I_{c2p}(c) \land c \in \sclasses _l \}$;\\
                }
                \If{$\spaths _r \neq \emptyset$}{ 
                    $\spaths _r \leftarrow \{(v,u) \mid (v,u) \in I_{c2p}(c) \land c \in \sclasses _r \}$;\\
                }
                $\spaths \leftarrow \{(v,u)\mid (v,u) \in \spaths _l \land (v,u) \in \spaths _r \land v=u\}$;\\
                {\bf return} $\spaths, \emptyset$;
              }
              \hrulefill \\
              {\bf procedure} {\sc Identity}($\spaths,  \sclasses$, $I_{k}$)\\
                \If{$\sclasses \neq \emptyset$}{
                    $\sclasses' \leftarrow \{c \mid c \in \sclasses \land (v,u) \in I_{c2p}(c) \land v=u\}$;\\
                    {\bf return} $\emptyset, \sclasses'$;
                }
                \Else{
                    $\spaths' \leftarrow \{(v,u) \mid (v,u) \in \spaths \land v=u\}$;\\
                    {\bf return} $\spaths', \emptyset$;
                    }
        }
\end{algorithm}

\vspace{0.1cm}
\begin{example}
Let us consider evaluating $\edgefont{ff} \cap \inverse{\edgefont{f}}$.
Our algorithm first {\sc LookUP} for $\edgefont{ff}$ and $\inverse{\edgefont{f}}$, and finds the sets of class identifiers $\{7, 16, 20\}$ and $\{7, 8, 9\}$, respectively.
Then, it takes {\sc Conjunction} for the two sets, and then obtains $\{7\}$.
Finally, we obtain three paths $\{(\mathit{sue}, \mathit{zoe}), (\mathit{joe}, \mathit{sue}), (\mathit{zoe}, \mathit{joe}) \}$ by $I_{c2p}(7)$.
Since we do not need to compare any paths for the conjunction operation, the query processing is very fast.

If we use the language-unaware path index, we first find 15 and 15 s-t pairs for $\edgefont{ff}$ and $\inverse{\edgefont{f}}$, resp., each element of which is a pair of vertex identifiers, i.e., in total, comparisons over 60 vertex identifiers.  
With \cpqindex, this would be an intersection of lists of class identifiers of length 3 and 3, resp., i.e., in total, comparisons over 6 class identifiers
This 10x reduction demonstrates the significant acceleration of \cpqindex.
\hfill{} $\square$
\end{example}

\noop{
\paragraph*{Correctness}
Query processing compares the class identifiers for evaluating {\sc Conjunction}.
Comparing class identifiers is equivalent to comparing paths.

\begin{proposition}
Given two sets of class identifiers $\shistories$ and $\sclasses'$, the set of paths $\spaths (c)$  for all $c \in \shistories \cap \shistories'$ is same as $\spathsch) \cap \spaths(h')$ for all $c \in \shistories$ and $c' \in \shistories'$.
\end{proposition}
\noindent {\it Proof:} The conjunction operation outputs paths that have the given two label sequences.
If class identifiers appear in both label sequences, the paths that are including in the class identifiers have both label sequences.
Thus, $\shistories \cap \shistories'$ is the set of class identifiers that appears for the given label sequences.
Therefore, $\spaths (h)$ for $h \in \shistories \cap \shistories'$ is the same as the set of paths that appear the sets of paths  $\shistories$ and $\shistories'$
\hfill{} $\square$
}

\subsection{Index maintenance}\label{ssec:maintenance}

\cpqindex{} is easily updated when the graph is updated.
Our update method lazily updates \cpqindex, while maintaining correctness of query evaluation. 
That is, to reduce update cost, it does not maintain the same index entries with the index that is constructed from scratch. 
This approach enables efficient index updates with a small deterioration of the index performance.

We proceed as follows.  When edges are deleted/inserted, label sequences of paths between s-t pairs change (also paths may disappear/appear) and $k$-path-bisimilar s-t pairs may become non-bisimilar.
If two non-bisimilar s-t pairs are assigned the same class identifier, then query results would be incorrect.
Thus, our lazy update method divides the set of paths for all $k$ length paths related to graph update and does not merge two sets of s-t pairs even if they become $k$-path-bisimilar. Query processing still ensures correct results even if $k$-path-bisimilar s-t pairs belong to the different class identifiers.
\begin{proposition}[Update correctness]\label{prop:updatecorrectness}
    After edge deletion or edge insertion, query processing in Sec.~\ref{ssec:query}
    ensures correct query results.
\end{proposition}

We explain how we handle five cases: edge deletion, edge insertion,  label change, vertex deletion, and vertex insertion. 

\noindent
{\bf Edge deletion.} 
We explain a procedure for edge deletion.
We first enumerate all s-t pairs involved in the deleted edge by bread-first search.
The label sequences of these s-t pairs may change unless there are alternative paths through same label sequences, so we check whether there are alternative paths.
Next, we delete paths from $I_{c2p}$ if the label sequences of the s-t pairs change. Here, for efficiently finding class identifier $c'$ according to the deleted s-t pairs, we use inverted index whose keys are s-t pairs. 
We then add new $\spaths(c')$ that includes only the path into $I_{c2p}$ unless their label sequences are empty (i.e., paths disappear).
This update does not check whether or not the affected s-t pairs is $k$-path-bisimilar to other s-t pairs.

\vspace{0.1cm}
\begin{example}
Suppose that we delete $\{(\mathit{ada}, \mathit{tim})\}$ with  $\edgefont{f}$ from $\graph_{ex}$ in Figure~\ref{fig:bisimulation_example}.
We first list all related s-t pairs with at most $2$ length such as $\{(\mathit{ada}, \mathit{123})\}$ and $\{(\mathit{tom}, \mathit{tim})\}$.
Among them, $\{(\mathit{ada}, \mathit{123})\}$ has alternative paths through $\edgefont{fv}$.
The other s-t pairs are deleted from corresponding blocks, and then new blocks with new class identifiers are created that include a single s-t pair for all deleted pairs.\hfill{} $\square$


\end{example}

\noindent
{\bf Edge insertion.} 
The procedure is similar to that for edge deletion.
The difference is enumerating s-t pairs involving the new inserted edge. 

\noindent
{\bf Other graph updates.} We can handle the following additional updates by combinations of edge deletion and insertion.
For example, in the vertex deletion, we delete all edges that connect to the deleted vertex, and then delete the vertex.

\noop{
We list how to handle other updates:
\begin{itemize}
    \item Label change: We delete the edge with current label then add the edge with new label.
    \item Vertex deletion: We delete all edges that connect to the deleted vertex.
    \item Vertex insertion: It does not affect to evaluation of the conjunctive path queries.
\end{itemize}
}





The update cost is much smaller than reconstructing the index from scratch.
After update, the set of $k$-path bisimilar s-t pairs may belong to different class identifiers.
We guarantee the correctness of query results even if the set of $k$-path bisimilar s-t pairs belong to different class identifiers.

\begin{theorem}
The time complexity for edge deletion or insertion is $O(d|\paths_u|+|\paths_u| \log|\paths^{\leq k}|+|\classes|\log|\classes|)$, 
where $\paths_u$ and $d$ are the set of s-t pairs that are involved updates and the maximum degrees among vertices in $\paths_u$, respectively.
\end{theorem}
{\it Proof}: The update method first finds $\paths_u$ by bread-first search starting from end vertices of an inserted/deleted edge. This takes $O(d|\paths_u|)$.
Then, it searches for s-t pairs in $\paths_u$ from $I_{c2p}$, and then it searches for the class identifiers associated with s-t pairs in $\paths_u$ from $I_{l2c}$. 
Since these data structures are sorted list, paths and class identifiers are found by two binary search, that is $O(|\paths_u|\log |\paths^{\leq k}| + |\classes| \log |\classes|)$.  
Therefore, the time complexity is $O(d|\paths_u|+|\paths_u| \log|\paths^{\leq k}|+|\classes|\log|\classes|)$.
\hfill{} $\square$

\noop{
\begin{proposition}[Update correctness]\label{prop:updatecorrectness}
    After edge deletion or edge insertion, query processing with Algorithm~\ref{alg:querywhole} ensures correct query results.
\end{proposition}

{\it Proof}: In the update, we separate the involved paths from class identifiers.
Let consider two paths that are $k$-path-bisimilar and whose class identifiers are different.
The algorithm for query processing handles two class identifiers separately, and if one class matches an expression, so does the other.
If not, another one also does not.
Therefore, the query processing answers the same set of paths before and after updates.   
\hfill{} $\square$
}

\section{Interest-Aware Index}
\label{sec:workload}

In many application scenarios, users are often interested in only a specific set of label sequences, i.e., navigation patterns.
We call the given label sequences {\it interests}.
Users are clearly interested in not all label sequences.
\cpqindex{}, however, stores all label sequences, including inverse of labels, up to length $k$, which leads to the scalability problem (i.e., large index construction costs and index size).

Motivated by this, we develop an interest-aware \cpqindex, namely \iacpqindex{} based on a given set of label sequences.
The \iacpqindex{} solves the scalability problem of \cpqindex.
The index supports processing of any \cpq, yet is tailored to especially accelerate processing of all \cpq{} queries which use any of the label sequences of interest.



\subsection{Index design}
Towards an interest-aware \cpqindex, we propose the notion of {\it interest-aware path-equivalence} as follows. 

\begin{definition}[Interest-aware Path-Equivalence] \label{def:workte}
Let \graph\ be a graph, 
$v, u, x, y\in \vertices$ and $\elabels_q \subseteq \elabels^{\leq k}$ be a set of label sequences. 
The s-t pairs $(v, u)$ and $(x, y)$ are {\em interest-aware path-equivalent}, denoted
$(v, u) \watransequibalent (x, y)$, if and only if the followings hold:
\begin{enumerate}
    \item $v=u$ if and only if $x=y$; 
    \item $\elabels^{\leq k}(v, u) \cap \elabels_q = \elabels^{\leq k}(x, y) \cap \elabels_q$.
\end{enumerate}
\end{definition} 

$\elabels_q$ is the set of interests. 
When we construct \iacpqindex, we always include all sequences of length one (i.e., all edge labels) in $\elabels_q$.
Thus, even queries containing label sequences without users' interests can be still evaluated. 


The difference between \cpqindex{} and \iacpqindex{} is that the former and the latter assign same class identifiers to the set of $k$-path bisimilar s-t pairs and the set of interest-aware path-equivalent s-t pairs, respectively.
Since interest-aware path-equivalence is weaker than $k$-path bisimulation (i.e., it is easy to show that $\kbisimilar$ refines $\watransequibalent$, when $k$ is at least as large as the length of the longest sequence in $\elabels_q$), more s-t pairs have the same class identifiers (i.e., partition blocks are bigger).
Therefore, the size of \iacpqindex{} is much smaller (and hence faster to use) than that of the basic \cpqindex, and we can control the size of index by adjusting the size of interests.

\begin{theorem}
    The size of \iacpqindex{} is $O\left(\frac{|\elabels_q|}{|\elabels^{\leq k}|} \left( \gamma | \classes |+ |\paths^{\leq k}|\right) \right)$.
\end{theorem}
{\it Proof:} The number of paths that are stored in $\iacpqindex$ linearly decreases by the size of $|\elabels_q|$, compared with the size of $\cpqindex$. Therefore, the size of \iacpqindex{} is $O\left(\frac{|\elabels_q|}{|\elabels^{\leq k}|} \left( \gamma | \classes |+ |\paths^{\leq k}|\right) \right)$.
\hfill{} $\square$

\subsection{Index construction and query processing}
The index construction and query processing methods are almost the same as those for \cpqindex. 
The difference for the construction algorithm is that it enumerates s-t pairs only with given label sequences and two paths have same class identifiers if they are interest-aware path-equivalent.
Since the construction of \iacpqindex{} decreases the number of paths, it becomes more efficient than that of \cpqindex.
The difference for query processing is that we divide label sequences into sub-label sequences if the label sequences are not included in the given label sequences.

The space and time complexity of constructing \iacpqindex{} are similar to Theorems \ref{thm:timecomplexity} and \ref{thm:spacecomplexity}, resp.
The construction cost of \iacpqindex{} decrease as $| \elabels_q |$ decreases since the number of paths related to its index construction decreases.

The time complexity of query processing on \iacpqindex{} is the same as that on \cpqindex. The difference is the numbers of paths and class identifiers by {\sc LookUp}. \iacpqindex{} reduces them, so the query processing becomes fast.



\smallskip

\subsection{Maintenance}
\iacpqindex{} is easily updated in a similar fashion for \cpqindex.


\noindent
{\bf Graph update:} The graph update procedures are almost the same as those for \cpqindex{} given in Section \ref{ssec:maintenance}.
The difference is that we do not process the set of s-t pairs whose label sequences are not included in the given set of label sequences.

\noindent
{\bf Label sequence deletion:} When we delete a label sequence from the given set of label sequences, we can just delete the deleted label sequence from $I_{l2c}$.
After deleting the label sequence, two paths may become interest-aware path-equivalent.
While we do not merge two sets of paths, we can still guarantee correct query answers in a fashion analogous with Proposition \ref{prop:updatecorrectness}.

\noindent
{\bf Label sequence insertion:} For inserting new label sequences, we insert new s-t pairs to the index.
Thus, we first enumerate the set of s-t pairs that have the inserted label sequences, and then take the same procedure as for inserting new edges.
\section{Experimental Study}
\label{sec:experiment}
We next present the results of an experimental evaluation of our methods.
We designed the experiments to clarify the questions:
(1) Does \CPQ-aware indexing
accelerate query processing?  (Section \ref{ssec:exp_efficiency});
(2) Are \CPQ-aware indexes compact? (Section \ref{ssec:exp_scalability}),
(3) Are \CPQ-aware indexes maintainable? (Section \ref{ssec:exp_update}); and,
(4) {\em Are CPQ-aware indexes well-behaved as $k$ grows?} (Section \ref{ssec:exp_analysis}).

\begin{figure*}[!th]
\scalebox{0.6}{
\captionsetup[subfigure]{labelformat=empty}
\begin{minipage}[t]{0.27\linewidth}
\subfloat[C2 $=\elabel_1 \cmp \elabel_2$]{
\begin{tikzpicture}[->,>=stealth',shorten >=1pt,auto,node distance=1.0cm,  thick,main node/.style={fill,circle,draw, minimum size=0.05cm,scale=0.5, font=\sffamily\footnotesize}]
  \draw[opacity=.0] (-0.5,0) rectangle (1.5,0);
  \node[main node] (1) at (0.5,-1.3)   {$ $};
  \node[main node] (2) [label=above:{$t$}] [right of=1] {$ $};
  \node[main node] (3) [label=above:{$s$}] [left of=1]  {$ $};
  \path[->]
    (1) edge [right] node[above] {$ $}   (2)
    (3) edge [left]  node[below] {$ $}      (1);
\end{tikzpicture}}
\subfloat[C4 $=$C2$\cmp$C2]{
\begin{tikzpicture}[->,>=stealth',shorten >=1pt,auto,node distance=1.0cm,  thick,main node/.style={fill,circle,draw, minimum size=0.05cm,scale=0.5,font=\sffamily\normalsize}]
  \node[main node] (1)  {$ $};
  \node[main node] (2)  [right of=1] [label=above:{$t$}] {$ $};
  \node[main node] (3) [left of=1]  {$ $};
  \node[main node] (4)  [left of=3] {$ $};
  \node[main node] (5)  [left of=4] [label=above:{$s$}] {$ $};
  \path[->, every node/.style={font=\small}]
    (1) edge [right] node[above] {$ $}   (2)
    (3) edge [left]  node[below] {$ $}   (1)
    (4) edge [right] node[above] {$ $}   (3)
    (5) edge [right] node[above] {$ $}   (4);
\end{tikzpicture}}
\centering
\end{minipage}
}
\scalebox{0.6}{
  \centering
  \captionsetup[subfigure]{labelformat=empty}
\begin{minipage}[t]{0.7\linewidth}
\subfloat[T $=$C2$\cap\elabel$]{ 
\begin{tikzpicture}[->,>=stealth',shorten >=1pt,auto,node distance=1.4cm,  thick,main node/.style={fill,circle,draw, minimum size=0.05cm,scale=0.5,font=\sffamily\normalsize}]
  \draw[opacity=.0] (-0.5,-1) rectangle (1.25,0);
  \node[main node] (1)  {$ $};
  \node[main node] (2)  [right of=1][label=below:{$t$}] {$ $};
  \node[main node] (3) [below of=1] [label=left:{$s$}] {$ $};
  \path[->, every node/.style={font=\small}]
    (1) edge [right] node[above] {$ $}   (2)
    (3) edge [right] node[above] {$ $}   (2)
    (3) edge [left]  node[below] {$ $}   (1);
\end{tikzpicture}}
\subfloat[S $=$C2$\cap$C2]{ 
\begin{tikzpicture}[->,>=stealth',shorten >=1pt,auto,node distance=1.4cm,  thick,main node/.style={fill,circle,draw, minimum size=0.05cm,scale=0.5,font=\sffamily\normalsize}]
  \draw[opacity=.0] (-0.5,-1) rectangle (1.25,0);
  \node[main node] (1)  {$ $};
  \node[main node] (2)  [right of=1][label=right:{$t$}] {$ $};
  \node[main node] (3) [below of=1] [label=left:{$s$}] {$ $};
  \node[main node] (4) [below of=2]  {$ $};
  \path[->, every node/.style={font=\small}]
    (1) edge [right] node[above] {$ $}   (2)
    (4) edge [right] node[above] {$ $}   (2)
    (3) edge [right] node[above] {$ $}   (4)
    (3) edge [left]  node[below] {$ $}      (1);
\end{tikzpicture}}
\subfloat[TT $=$T$\cap$C2]{ 
\begin{tikzpicture}[->,>=stealth',shorten >=1pt,auto,node distance=1.4cm,  thick,main node/.style={fill,circle,draw, minimum size=0.05cm,scale=0.5,font=\sffamily\normalsize}]
  \draw[opacity=.0] (-1.0,-1) rectangle (1.5,0);
  \node[main node] (1)  {$ $};
  \node[main node] (2)  [right of=1][label=right:{$t$}] {$ $};
  \node[main node] (3) [below of=1] [label=left:{$s$}] {$ $};
  \node[main node] (4) [below of=2]  {$ $};
  \path[->, every node/.style={font=\small}]
    (1) edge [right] node[above] {$ $}   (2)
    (4) edge [right] node[above] {$ $}   (2)
    (3) edge [right] node[above] {$ $}   (2)
    (3) edge [right] node[above] {$ $}   (4)
    (3) edge [left]  node[below] {$ $}      (1);
\end{tikzpicture}}
\subfloat[TC $=$T$\cmp\elabel$]{ 
\begin{tikzpicture}[->,>=stealth',shorten >=1pt,auto,node distance=1.4cm,  thick,main node/.style={fill,circle,draw, minimum size=0.05cm,scale=0.5,font=\sffamily\normalsize}]
  \draw[opacity=.0] (-0.5,-1) rectangle (1.25,0);
  \node[main node] (1)  {$ $};
  \node[main node] (2)  [right of=1] {$ $};
  \node[main node] (3) [below of=1] [label=left:{$s$}] {$ $};
  \node[main node] (4) [below of=2] [label=right:{$t$}] {$ $};
  \path[->, every node/.style={font=\small}]
    (1) edge [right] node[above] {$ $}   (2)
    (2) edge [right] node[above] {$ $}   (4)
    (3) edge [right] node[above] {$ $}   (2)
    (3) edge [left]  node[below] {$ $}      (1);
\end{tikzpicture}}~\quad
\subfloat[SC $=$S$\cmp \elabel$]{
\begin{tikzpicture}[->,>=stealth',shorten >=1pt,auto,node distance=1.4cm,  thick,main node/.style={fill,circle,draw, minimum size=0.05cm,scale=0.5,font=\sffamily\normalsize}]
  \draw[opacity=.0] (-0.5,-1) rectangle (1.25,0);
  \node[main node] (1)  {$ $};
  \node[main node] (2)  [right of=1] {$ $};
  \node[main node] (3) [below of=1] [label=left:{$s$}]  {$ $};
  \node[main node] (4) [below of=2]  {$ $};
  \node[main node] (5) [right of=2] [label=below:{$t$}] {$ $};
  \path[->, every node/.style={font=\small}]
    (1) edge [right] node[above] {$ $}   (2)
    (4) edge [right] node[above] {$ $}   (2)
    (2) edge [right] node[above] {$ $}   (5)
    (3) edge [right] node[above] {$ $}   (4)
    (3) edge [left]  node[below] {$ $}      (1);
\end{tikzpicture}}
\subfloat[ST $=$S$\cmp$T]{
\begin{tikzpicture}[->,>=stealth',shorten >=1pt,auto,node distance=1.4cm,  thick,main node/.style={fill,circle,draw, minimum size=0.05cm,scale=0.5,font=\sffamily\normalsize}]
  \draw[opacity=.0] (-0.5,-1) rectangle (1.25,0);
  \node[main node] (1)  {$ $};
  \node[main node] (2)  [right of=1] {$ $};
  \node[main node] (3) [below of=1] [label=left:{$s$}] {$ $};
  \node[main node] (4) [below of=2]  {$ $};
  \node[main node] (5) [right of=2]  {$ $};
  \node[main node] (6) [below of=5] [label=left:{$t$}] {$ $};
  \path[->, every node/.style={font=\small}]
    (1) edge [right] node[above] {$ $}   (2)
    (4) edge [right] node[above] {$ $}   (2)
    (2) edge [right] node[above] {$ $}   (5)
    (3) edge [right] node[above] {$ $}   (4)
    (5) edge [right] node[above] {$ $}   (6)
    (2) edge [right] node[above] {$ $}   (6)
    (3) edge [left]  node[below] {$ $}   (1);
\end{tikzpicture}}
\centering
\end{minipage}
}
\scalebox{0.6}{
\captionsetup[subfigure]{labelformat=empty}
\begin{minipage}[t]{0.65\linewidth}
\subfloat[C2i $=$C2$\cap$id]{
\begin{tikzpicture}[->,>=stealth',shorten >=1pt,auto,node distance=1.8cm,  thick,main node/.style={fill,circle,draw, minimum size=0.05cm,scale=0.5, font=\sffamily\footnotesize}]
  \draw[opacity=.0] (-0.5,0) rectangle (1.5,0);
  \node[main node] (1) [label=above:{$s,t$}]  {$ $};
  \node[main node] (2) [right of=1] {$ $};
  \path[->]
    (1) edge [bend left] node[above] {$ $}   (2)
    (2) edge [bend left]  node[below] {$ $}      (1);
\end{tikzpicture}}~\quad
\subfloat[Ti $=$(C2$\cmp$\elabel)$\cap$id]{
\begin{tikzpicture}[->,>=stealth',shorten >=1pt,auto,node distance=1.4cm,  thick,main node/.style={fill,circle,draw, minimum size=0.05cm,scale=0.5,font=\sffamily\normalsize}]
  \draw[opacity=.0] (-0.5,0) rectangle (1.5,0);
  \node[main node] (1)  {$ $};
  \node[main node] (2)  [right of=1] {$ $};
  \node[main node] (3) [below of=1][label=left:{$s,t$}]  {$ $};
  \path[->, every node/.style={font=\small}]
    (1) edge [right] node[above] {$ $}   (2)
    (2) edge [right] node[above] {$ $}   (3)
    (3) edge [left]  node[below] {$ $}      (1);
\end{tikzpicture}}~\quad
\subfloat[Si $=$C4$\cap$id]{
\begin{tikzpicture}[->,>=stealth',shorten >=1pt,auto,node distance=1.4cm,  thick,main node/.style={fill,circle,draw, minimum size=0.05cm,scale=0.5,font=\sffamily\normalsize}]
  \draw[opacity=.0] (-0.5,-1) rectangle (1.25,0);
  \node[main node] (1)  {$ $};
  \node[main node] (2)  [right of=1] {$ $};
  \node[main node] (3) [below of=1][label=left:{$s,t$}]  {$ $};
  \node[main node] (4) [below of=2]  {$ $};
  \path[->, every node/.style={font=\small}]
    (1) edge [right] node[above] {$ $}   (2)
    (2) edge [right] node[above] {$ $}   (4)
    (4) edge [right] node[above] {$ $}   (3)
    (3) edge [left]  node[below] {$ $}      (1);
\end{tikzpicture}}
\subfloat[St $=$($\elabel_1\cmp\inverse{\elabel_1}$)$\cap$($\elabel_2\cmp\inverse{\elabel_2}$)$\cap$($\elabel_3\cap\inverse{\elabel_3}$)$\cap$id]{
\begin{tikzpicture}[->,>=stealth',shorten >=1pt,auto,node distance=1.4cm,  thick,main node/.style={fill,circle,draw, minimum size=0.05cm,scale=0.5,font=\sffamily\normalsize}]
  \draw[opacity=.0] (-0.5,-1) rectangle (3,0);
  \node[main node] (1) [label=left:{$s,t$}] {$ $};
  \node[main node] (2)  [right of=1] {$ $};
  \node[main node] (3) [below of=1]  {$ $};
  \node[main node] (4) [below of=2]  {$ $};
  \path[->, every node/.style={font=\small}]
    (1) edge [right] node[above] {$ $}   (2)
    (1) edge [right] node[above] {$ $}   (4)
    (1) edge [left]  node[below] {$ $}   (3);
\end{tikzpicture}}
\centering
\end{minipage}
}
\vspace{-3mm}
\caption{Query templates, where $s$ and $t$ denote the source and target of paths, respectively. } 
\vspace{-3mm}
\label{fig:querytemplates}
\end{figure*}
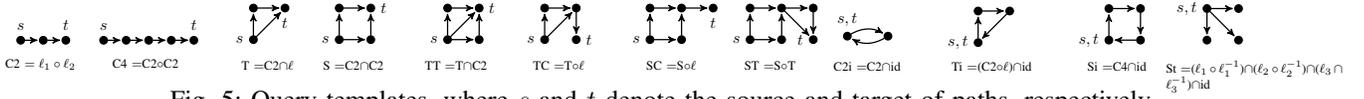

\begin{table}[]
    \centering
    \caption{Dataset overview: $|\edges|$ and $|\elabels|$ include inverse edges and labels, respectively.}
    \label{tab:dataset}
    {\small
    \begin{tabular}{|c|r|r|r|c|}\hline
         Dataset  &\multicolumn{1}{|c|}{$|\vertices|$} & \multicolumn{1}{|c|}{$|\edges|$} & \multicolumn{1}{|c|}{$|\elabels|$}&Real label?\\ \Hline
         {\bf Robots} & 1{,}484 & 5{,}920 & 8&\checkmark\\
         {\bf ego-Facebook} & 4{,}039 & 176{,}468 & 16& \\
         {\bf Advogato} & 5{,}417 & 102{,}654 & 8&\checkmark \\         
         {\bf Youtube}& 15{,}088 & 21{,}452{,}214 & 10&\checkmark \\    
         {\bf StringHS} & 16{,}956 & 2{,}483{,}530 & 14&\checkmark\\  
         {\bf StringFC} & 15{,}515 & 4{,}089{,}600 & 14&\checkmark \\         
         {\bf BioGrid} & 64{,}332 & 1{,}724{,}554 & 14&\checkmark\\              
         {\bf Epnions} & 131{,}828 & 1{,}681{,}598 & 16& \\
         {\bf WebGoogle}& 875{,}713 & 10{,}210{,}074 & 16&  \\     
         {\bf WikiTalk}& 2{,}394{,}385 & 10{,}042{,}820& 16&  \\      
         {\bf YAGO}& 4{,}295{,}825 & 24{,}861{,}400 & 74&\checkmark\\
         {\bf CitPatents}& 3{,}774{,}768 & 33{,}037{,}896 & 16& \\ 
         {\bf Wikidata}& 9{,}292{,}714 & 110{,}851{,}582 & 1054&\checkmark \\
         {\bf Freebase}& 14{,}420{,}276 & 213{,}225{,}620 & 1556 &\checkmark \\\hline 
         {\bf g-Mark-1m} & 1{,}006{,}802 & 15{,}925{,}506 & 12& \\
         {\bf g-Mark-5m} & 5{,}005{,}992 & 84{,}994{,}500 & 12& \\
         {\bf g-Mark-10m} & 10{,}005{,}721 & 183{,}748{,}319 & 12& \\
         {\bf g-Mark-15m} & 15{,}003{,}647 & 255{,}538{,}724 & 12& \\
         {\bf g-Mark-20m} & 20{,}004{,}856 & 393{,}797{,}046 & 12& \\\hline
    \end{tabular}
    }
\end{table}

Experiments were performed on a Linux server with 512GB of memory and an Intel(R) Xeon(R) CPU E5-2699v3 @ 2.30GHz processor. All algorithms are single-threaded.

\smallskip
\noindent {\bf Datasets.} 
Table \ref{tab:dataset} provides an overview of the datasets used in our study consisting of nine datasets with real labels and five datasets without edge labels.
The graphs range over several different scenarios, such as social networks, biological networks, and knowledge graphs.
These datasets are provided by the authors~\cite{Valstar2017,peng2020answering} except for ego-Facebook, webGoogle, WebTalk, CitPatents, and Wikidata. 
In Wikidata, we extract vertices that represent URI from original Wikidata~\cite{Wikidata}.
ego-Facebook, WebGoogle, WebTalk, and CitPatents are available at SNAP~\cite{snap}.
Since these four graphs have no edge labels, we assign edge labels that are exponentially distributed with $\lambda = 0.5$ which follows the distribution of edge labels on YAGO.
Note that the graphs are of the same size and complexity as those used in recent studies \cite{han2019efficient,Valstar2017,peng2020answering,sun2020rapidmatch}.

The synthetic datasets model citation networks with three types of vertices, {\edgefont researcher}, {\edgefont venue}, and {\edgefont city}, and six edge labels, {\edgefont cites} from/to researchers, {\edgefont supervises} from/to researchers, {\edgefont livesIn} from researcher to city, {\edgefont workisIn} from researcher to city, {\edgefont publishesIn} from researcher to venue, and {\edgefont heldIn} from venue to city.
We use the synthetic datasets for evaluating scalability, varying the number of vertices and edges from roughly 1 and 8 million ({\bf g-Mark-1m}) to 20 and 200 million ({\bf g-Mark-20m}), resp.

\smallskip
\noindent {\bf Queries.} 
We used twelve \CPQ{} templates as described in Figure~\ref{fig:querytemplates}.
These query structures correspond to practical structures appearing, e.g., in the Wikidata query logs \cite{bonifati2020analytical, BonifatiMT19}.
We use as abbreviations C, T, S, and St for Chain, Triangle, Square, and Star, respectively. 
We especially chose these templates to (1) better understand the interaction of all basic constructs of the language and (2) exemplify query structures occurring frequently in practice such as chains (e.g., C4), stars (St), cycles (e.g., Ti), and flowers (e.g., ST).




For each template and dataset, we generate ten queries with random labels.
We only use queries in which all (sub-)paths of length two are non-empty, but the answers of some queries may be empty.
To evaluate the difference between query times of non-empty and empty queries, queries on Yago, Wikidata, and Freebase have 50$\%$ non-empty and 50$\%$ empty queries except for C2. Queries for other datasets consist of mostly non-empty queries though we randomly set labels.
We report for each query template the average response time over all ten queries.
Note that the answers of some queries may be empty, but intermediate results are non-empty.

\smallskip
\noindent {\bf Methods.} We compare the following methods: 
    {\bf \cpqindex}, our \CPQ-aware index of Section \ref{sec:index};
    {\bf \iacpqindex}, our interest-aware \cpqindex{} of Section \ref{sec:workload};
    {\bf Path}, the state-of-the-art lungauge-unaware path index proposed in \cite{fletcher2016efficient}; 
    {\bf iaPath}, {\bf Path} where only label sequences included in the given interest are indexed; 
    {\bf TurboHom++}, the state-of-the-art algorithm for homomorphic subgraph matching \cite{kim2015taming}\footnote{The binary code of TurboHom++ was provided by the authors \cite{kim2015taming}.}; 
    {\bf Tentris}, the state-of-the-art RDF engine~\cite{bigerl2020tentris}\footnote{\url{https://github.com/dice-group/tentris}}; and,
    {\bf BFS}, index-free breadth-first-search query evaluation \cite{Bonifati2018}.
    We implemented all methods (see our open source codebase) except for TurboHom++ and Tentris.
    To be fair, we used the same query plans for all methods, except for TurboHom++ and Tentris which perform their own planning. 

    A {\bf relational database approach} is essentially the same as Path with $k=1$, which has lower performance than with $k=2$.
    {\bf RDF3X} and {\bf Virtuoso} were shown to be outperformed by TurboHom++~\cite{kim2015taming} and Tentris~\cite{bigerl2020tentris}.
    Thus, we exclude Virtuoso, RDF3X, and the relational graph approach in our experiments.

We varied path length $k$ from one to four, with a default value of two. 
For the interest-aware indexes on the datasets, we specify all label sequences in the set of queries as the interests. We divide label sequences larger than $k$ length into prefix label sequences of length $k$ and the rest.
On synthetic datasets, we specify five label sequences as interests; {\edgefont cites}-{\edgefont cites}, {\edgefont cites}-{\edgefont supervises}, {\edgefont publishesIn}-{\edgefont heldIn}, {\edgefont worksIn}-{\edgefont heldIn}$^{-1}$, and {\edgefont livesIn}-{\edgefont worksIn}$^{-1}$.

\smallskip

\noindent
{\bf Index implementation.} 
In this study we use simple in-memory data structures;
the study of alternative physical index representations is an interesting topic beyond the scope of this paper.
Identifiers of vertices and labels are 32-bit integers, following TurboHom++. 
Indexes are implemented as standard C++ vectors. For further details, please see our open-source codebase.





\begin{figure*}[ttt]
\centering
\begin{minipage}[t]{1.0\linewidth}
    \centering
    \includegraphics[width=1.0\linewidth]{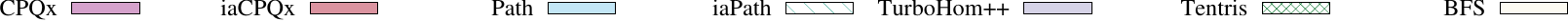}
    \end{minipage}
    \\
\begin{minipage}[t]{0.91\linewidth}
    \centering
	\includegraphics[width=1.0\linewidth]{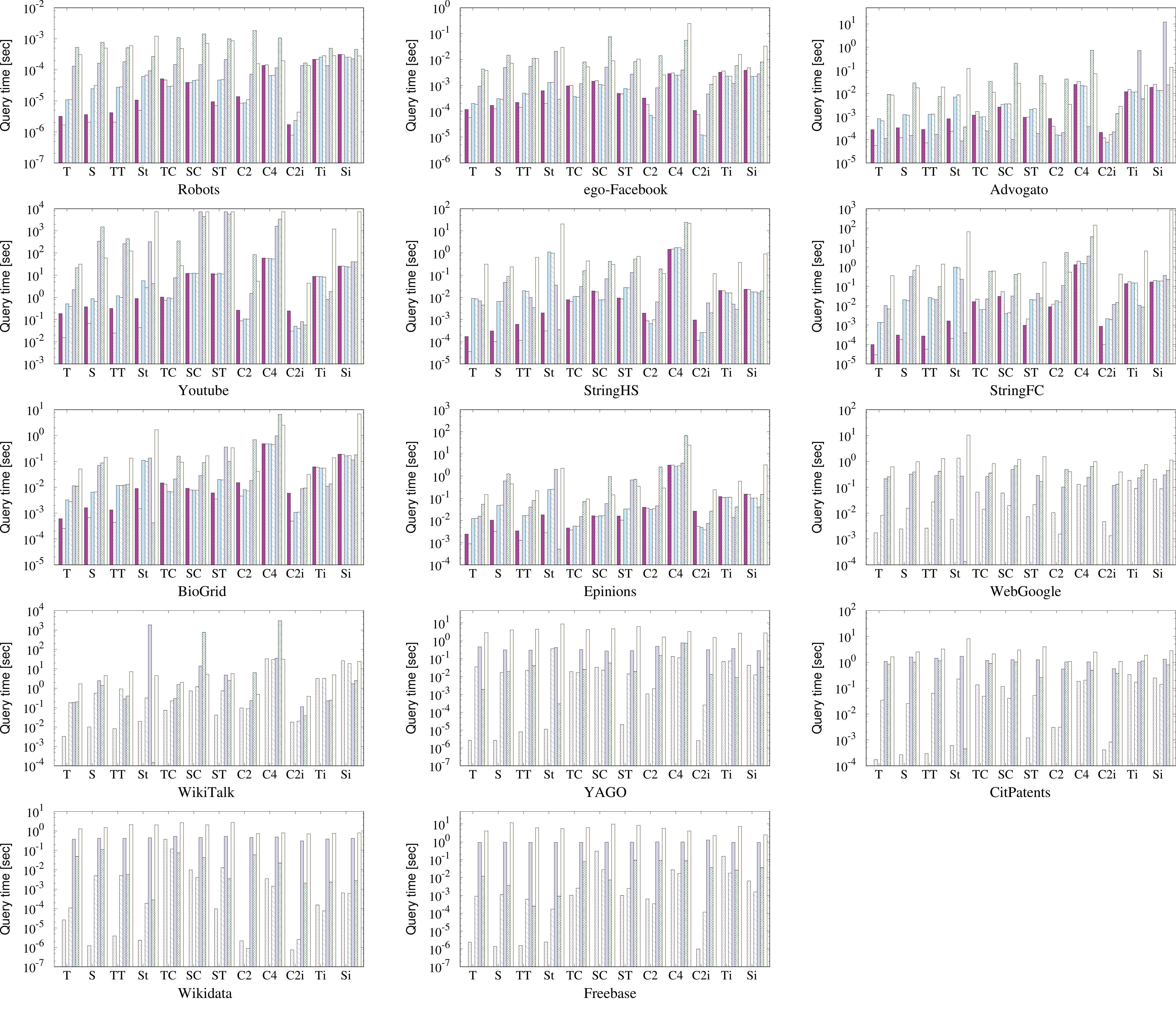}
	\vspace{-5 mm}
	\caption{Average query time for 12 query templates on real datasets. We terminated queries if the queries did not finish within two hours. 
    Note that \cpqindex{} and Path are not reported for WebGoogle, WikiTalk, YAGO, CitPatents, Wikidata, and Freebase due to out of memory.}
	\label{fig:querytime}
	\end{minipage}
\end{figure*}

\begin{figure*}[ttt]
\begin{tabular}{cc}
  \begin{minipage}{.39\textwidth}
  \centering
\tblcaption{The numbers of class identifiers on \cpqindex{} and \iacpqindex{} and the number of s-t pairs on iaPath, for evaluating S queries. 
}
\label{tab:numhistory}
{\footnotesize
\begin{tabular}{|c|r|r|r|}\hline
      \multicolumn{1}{|c|}{Dataset}&\multicolumn{1}{|c|}{\cpqindex} &\multicolumn{1}{|c}{\iacpqindex}&\multicolumn{1}{|c|}{iaPath}\\ \Hline
     {\bf Robots} & 0.4K & 0.13K  &  2.4K \\
     {\bf ego-Facebook} &22K & 19K &  23K \\
     {\bf Youtube} & 18M & 2.0M &  21M\\
     {\bf Epinions} &715K & 222K  &  1.8M \\
     {\bf Advogato} & 38.0K & 6.4K &  93.6K \\
     {\bf BioGrid} & 275K & 71.2K  &  499K \\         
     {\bf StringHS} & 49.7K & 11.2K&  750K \\        
     {\bf StringFC} & 33.3K & 3.7K&  388K\\
     {\bf Yago} & - & 75 &  967K\\     
     {\bf WikiTalk} &- & 915K &  19M\\  
     {\bf WebGoogle} & - & 287K &  760K\\  
     {\bf CitPatents} & - & 36K &   1.1M \\
     {\bf Wikidata} & - & 3 & 286M\\ 
     {\bf Freebase} & - & 8.6 &   79K \\
     \hline
\end{tabular}
}
  \end{minipage}
  \quad
  \begin{minipage}{.55\textwidth}
  \begin{minipage}[t]{1.0\linewidth}
    \centering
    \includegraphics[width=1.0\linewidth]{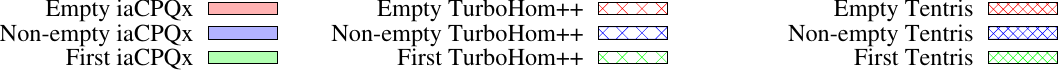}
\end{minipage}
\\
\begin{minipage}[t]{1.0\linewidth}
     \centering
     \vspace{-3mm}
  \subfloat[Yago]{\epsfig{file=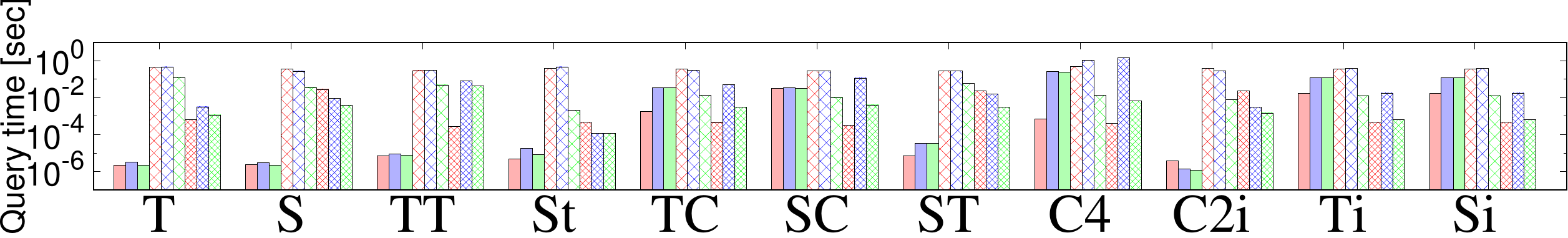,width=1.0\linewidth}
  }\\
       \vspace{-3mm}
  \subfloat[Wikidata]{\epsfig{file=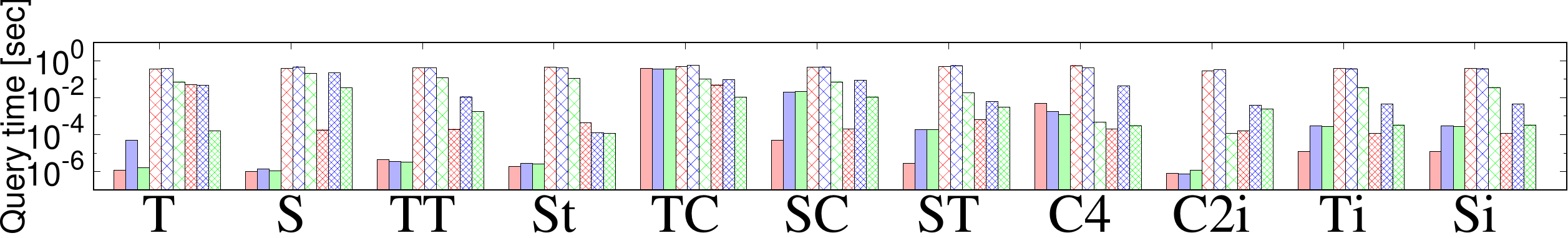,width=1.0\linewidth}
  }\\
       \vspace{-3mm}
\subfloat[Freebase]{\epsfig{file=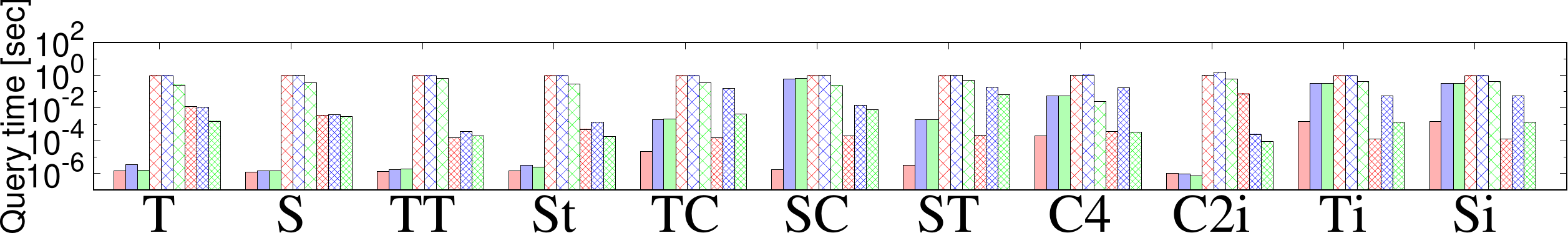,width=1.0\linewidth}
  }
\end{minipage}
\vspace{-2mm}
  \caption{Average query time of empty and non-empty queries, and for obtaining the first result of non-empty queries}
 \label{fig:negaposi}
  \end{minipage}
\end{tabular}
\vspace{-2mm}
\end{figure*}

\subsection{Does CPQ-aware indexing accelerate query processing? }
\label{ssec:exp_efficiency}

In summary, we can answer ``yes'' for this question.
Figure \ref{fig:querytime} shows the average query time of each method for each of the twelve query templates on nine datasets.
We do not show results on other five datasets due to page limitation.

\CPQ-aware indexes accelerate {\sc Conjunction} as mentioned at Section \ref{ssec:query}, so query times of T, S, TT, and St with \cpqindex{} and \iacpqindex{} are significantly smaller than those with all methods.
For TC, SC, and ST, either \iacpqindex{} or Path are the fastest on many datasets.
When {\sc Conjunction} is heavy, \iacpqindex{} is advantageous. 
For queries with {\sc Join} and without {\sc Conjunction} such as C2, C4, Ti, and Si, since \CPQ-aware indexes take two accesses to both $I_{l2c}$ and $I_{c2p}$, they have higher costs than Path, but the difference between them is small.
Query time of C2i is smaller than that of C2 in both \iacpqindex{} and Path.
This is because the size of answers decreases, and thus a cost for inserting s-t pairs to the answer sets reduces.
Efficient {\sc Identity} works well on some datasets such as Robots, YAGO, and Freebase while the efficiency highly depends on specified labels.
For Ti and Si, TurboHom++ and Tentris perform well on some datasets because it joins only s-t pairs that satisfy cycle but other methods check whether s-t pairs are cycle or not after join. Tentris additionally performs well for St query on some dataset due to its query optimization.
Compared with TurboHom++ and Tentris, our methods have significant improvement for many query templates such as T, S, TT, St, C2, and C2i. In particular, TurboHom++ and Tentris do no finish some queries within two hours in experiments; whereas \iacpqindex{} and Path finish all queries within two hours.

Comparing \cpqindex{} with \iacpqindex, \iacpqindex{} is more efficient because the numbers of class identifiers is smaller.
In particular, for C2i, \iacpqindex{} is much faster than \cpqindex{} as it reduces the number of {\sc LookUp}. 
Here, we note that iaPath does not become faster than Path because both of the indexes have the same number of s-t pairs regarding to label sequences.

We evaluate synthetic benchmarking queries for three datasets, YAGO2~\cite{yago2}, LUBM~\cite{lubm}, and WatDiv~\cite{watdiv}.
We use Y1--Y4 queries generated in \cite{harbi2016accelerating} for YAGO2, L1--L7 queries for LUBM and L1--L5 and S1--S7 queries for WatDiv.
For these benchmarking queries, we transform them into CPQs with keeping query shapes and their edge labels. We assign source and targets to the queries by ourselves.


\smallskip
\noindent
{\bf Pruning power.} We show the reason why our \CPQ-aware indexes accelerate query processing more than the-state-of-the-art language-unaware path indexes. 
Recall that our query processing algorithm accelerates queries with {\sc Conjunction} by comparing class identifiers instead of s-t pairs. 
Table \ref{tab:numhistory} shows the average numbers of class identifiers in \cpqindex{} and \iacpqindex{} and the average number of s-t pairs in iaPath, which are involved on evaluating S queries. 
The smaller numbers indicate higher pruning power. 

The numbers of class identifiers that are involved during the query evaluation in \cpqindex{} and \iacpqindex{} are much smaller than the number of s-t pairs in iaPath.
This result shows that partitioning s-t pairs based on $k$-path-bisimulation is effective for evaluating \CPQ.

\smallskip
\noindent
{\bf Impact of empty and non-empty queries.} We evaluate the impact of empty and non-empty results on query time.
The purposes here are (1) to gain further insight into the performance of our methods and (2) to compare the search strategy of our algorithm with that of TurboHom++ and Tentris. 
Figure \ref{fig:negaposi} shows the query time on empty and non-empty queries on Yago, Wikidata, and Freebase of \iacpqindex{}, TurboHom++, and Tentris. 
TurboHom++ outputs subgraphs, whereas \CPQ s have binary output (i.e., s-t pairs). 
For a closer comparison, we also evaluate the query time for finding the first answer (thereby offsetting the cost of enumerating all non-binary answers with TurboHom++), also shown in Figure \ref{fig:negaposi}.

From these results, we can see that \iacpqindex{} is much faster than TurboHom++ and Tentris for both empty and non-empty queries on most datasets and query templates.
The query time on empty queries is generally smaller than that on non-empty queries because (1) empty queries do not have insert cost to the answers and (2) empty queries might terminate on the way of query evaluation due to empty intermediate results.
Some non-empty queries are faster than empty queries when the intermediate results on empty queries are large.

\begin{figure}[ttt]
\begin{minipage}[t]{1.0\linewidth}
    \centering
    \includegraphics[width=1.0\linewidth]{figures/experiment/graphsize_query_legends.pdf}
\end{minipage}
\\
\begin{minipage}[t]{1.0\linewidth}
     \centering
    \includegraphics[width=0.95\linewidth]{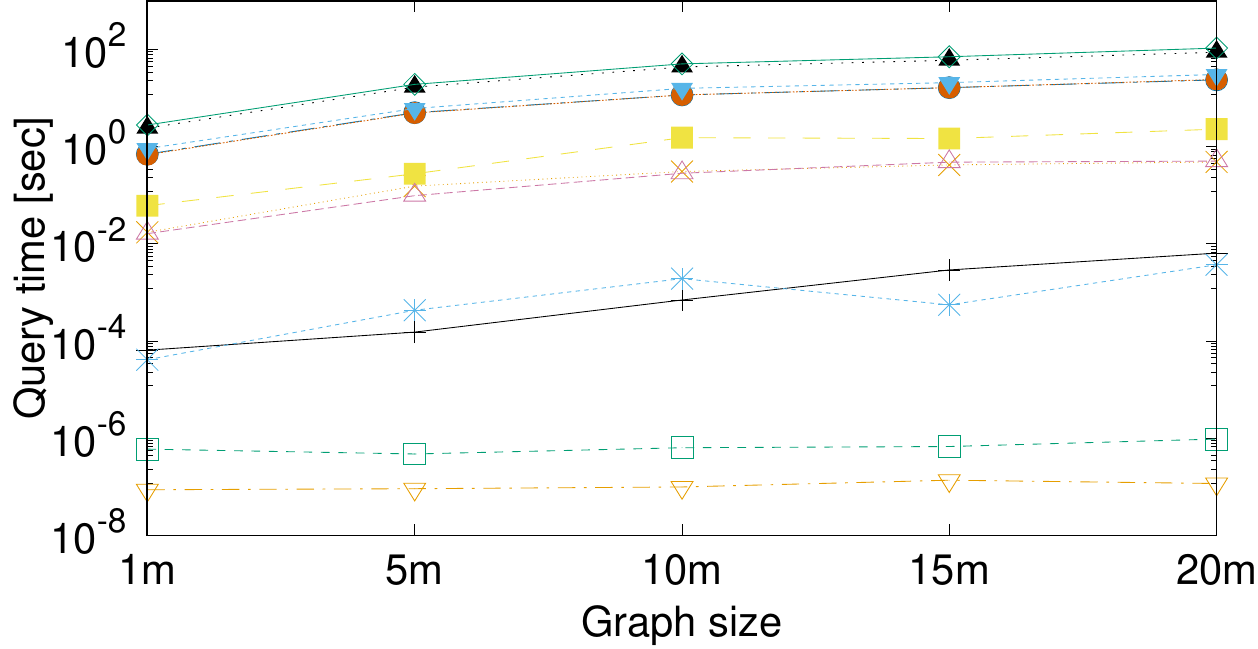}
\end{minipage}
\caption{Query performance of \iacpqindex{} as graph size grows}
\label{fig:query_graphsize}
\end{figure}

{\bf Scalability:} Figure \ref{fig:query_graphsize} shows the average query time of \iacpqindex{} for varying graph size on synthetic datasets. Our method scalably evaluates \CPQ s as graphs grow larger.

\begin{figure*}[ttt]
\begin{minipage}{.4\textwidth}
\begin{minipage}[t]{1.0\linewidth}
    \centering
    \includegraphics[width=1.0\linewidth]{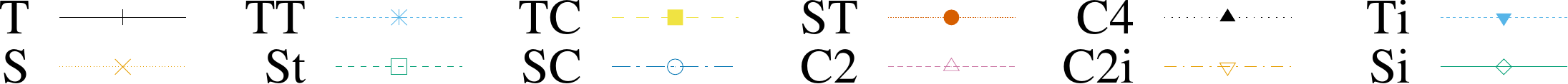}
\end{minipage}
\\
\begin{minipage}[t]{1.0\linewidth}
     \centering
    \includegraphics[width=0.95\linewidth]{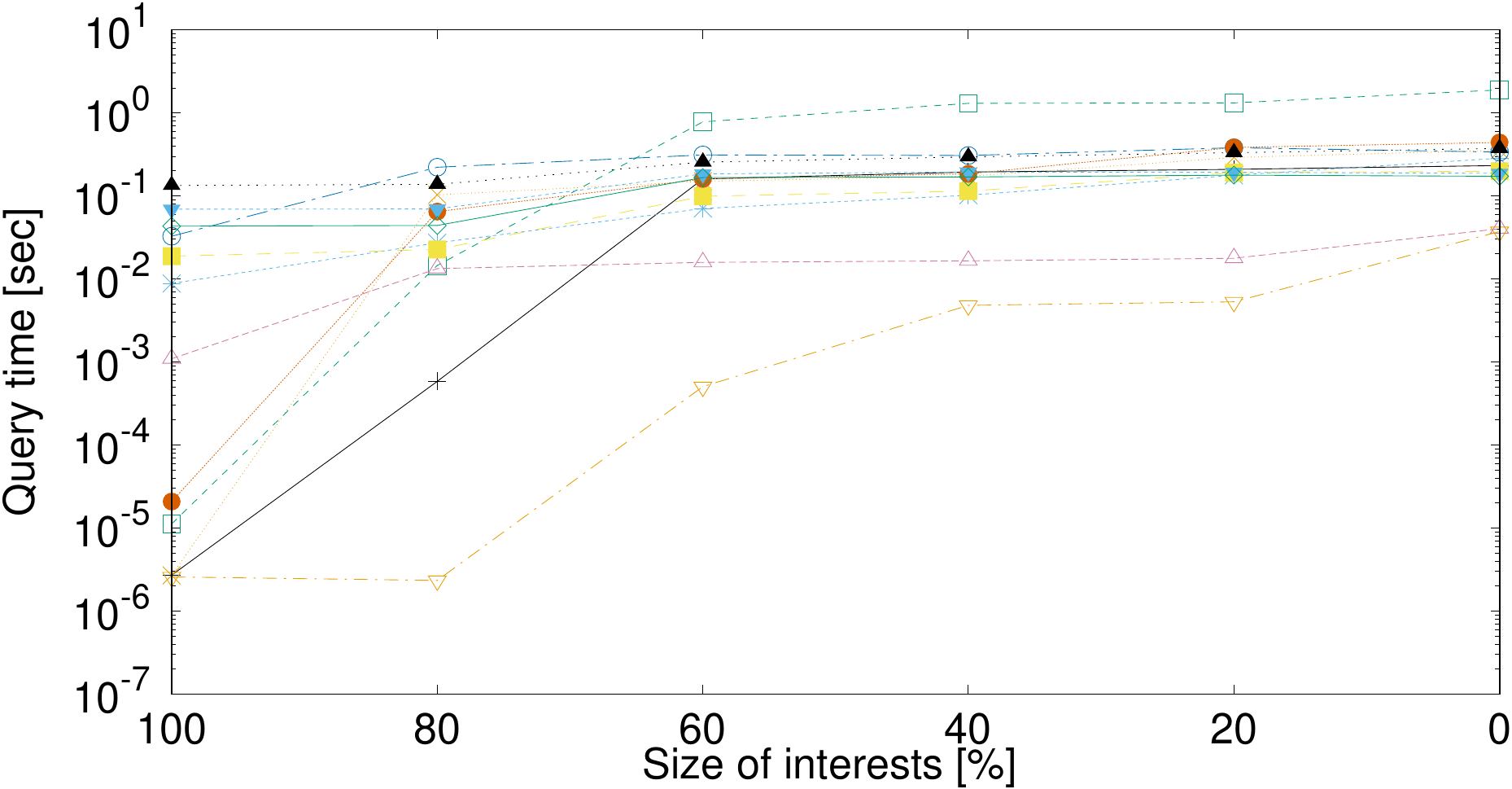}
\end{minipage}
\caption{Impact of interest size to query time \iacpqindex{} on YAGO. Values on X-axis indicate the percentage of label sequences in the set of queries that we use as interests.}
\label{fig:query_workloadchange}
\end{minipage}
\begin{minipage}{.27\textwidth}
    \centering
	\includegraphics[width=1.0\linewidth]{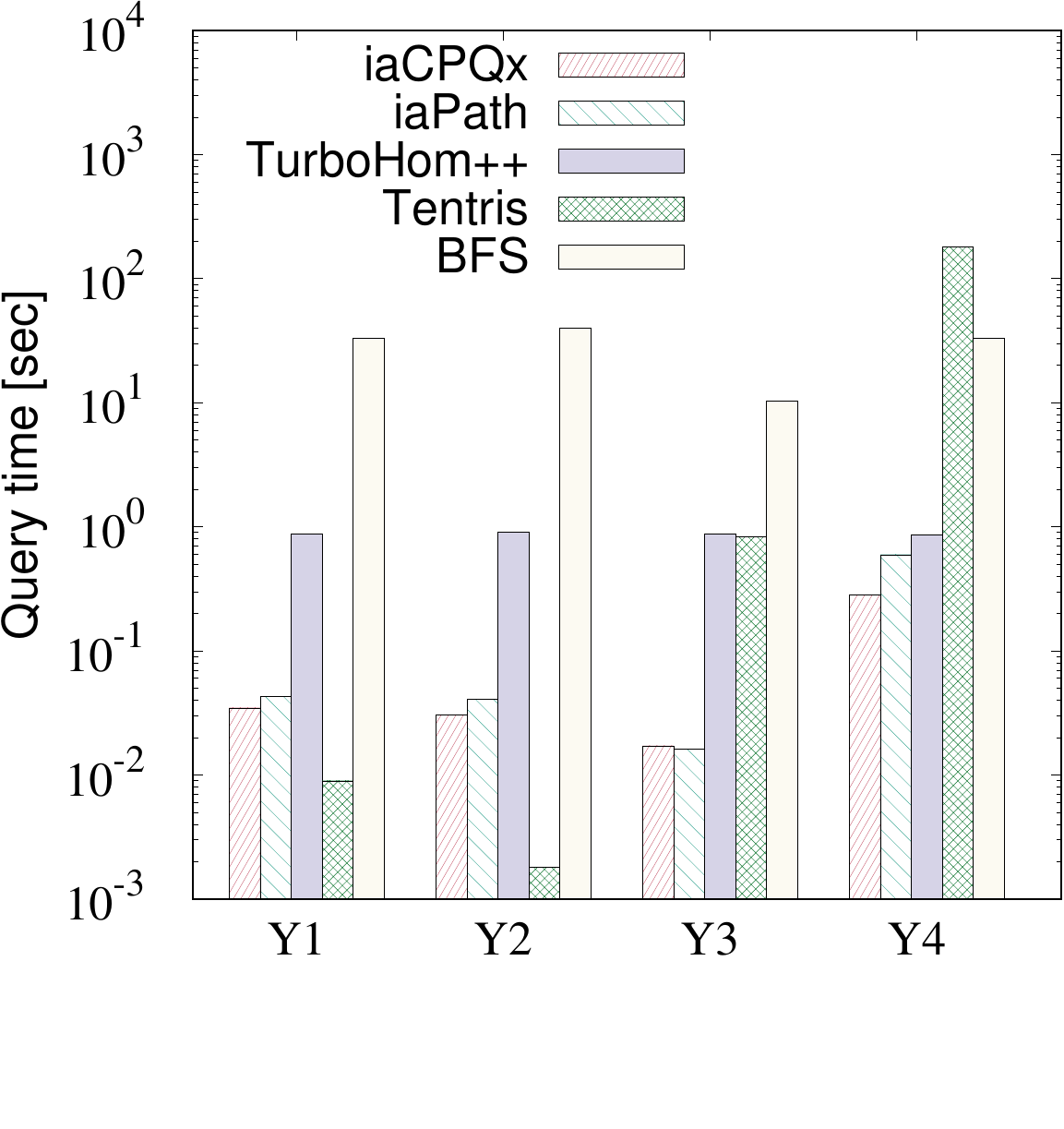}
	\vspace{-13mm}
    \caption{Query time on YAGO2 benchmarking queries}
    \label{fig:yago2}
    \vspace{-5mm}
\end{minipage}
\begin{minipage}{.25\textwidth}
    \centering
	\includegraphics[width=1.0\linewidth]{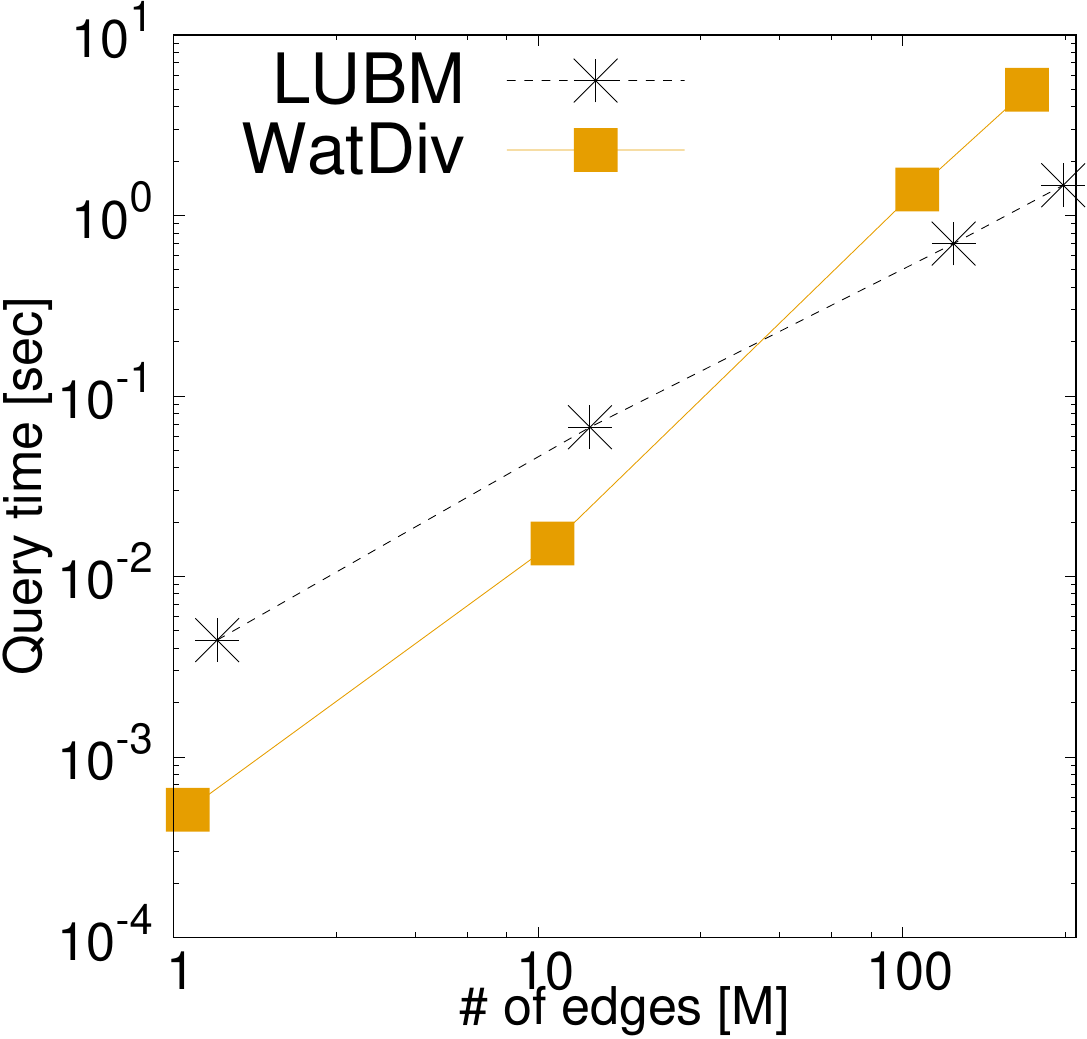}
    \caption{Query time on LUBM and WatDiv}
    \label{fig:lubmwatdiv}
    \vspace{-5mm}
\end{minipage}
\end{figure*}


Figure~\ref{fig:yago2} shows the query time on YAGO benchmarking queries.
YAGO2 includes 80 M vertices, 164 M edges, and 38 edge labels (including inverse edge labels).
\iacpqindex{} averagely achieves the smallest query time among them.
Figure~\ref{fig:lubmwatdiv} shows the average query time of given queries varying with graph sizes of LUBM and WatDiv. \iacpqindex{} can be built on graphs with about 280 and 220 M edges for LUBM and WatDiv, respectively. 
The increasing ratio of query time depends on query shapes. WatDiv benchmarking query needs more joins than ones of LUBM, so its query time largely increases as graph sizes increase.

\begin{table*}[ttt]
    \centering
    \caption{Index size (IS) and index construction time (IT), where ``-'' indicates out of memory.}
    \label{tab:index}
    {\footnotesize
    \begin{tabular}{|c|r|r|r|r|r|r|r|r|}\hline
         \multirow{2}{*}{Dataset}& \multicolumn{2}{|c|}{\cpqindex}&\multicolumn{2}{|c|}{\iacpqindex}&\multicolumn{2}{|c|}{Path}&\multicolumn{2}{|c|}{iaPath}  \\
          &\multicolumn{1}{c}{IS [B]}&\multicolumn{1}{c|}{IT [s]}&\multicolumn{1}{c}{IS [B]}&\multicolumn{1}{c|}{IT [s]}&\multicolumn{1}{|c}{IS [B]}&\multicolumn{1}{c|}{IT [s]}&\multicolumn{1}{|c}{IS [B]}&\multicolumn{1}{c|}{IT [s]}\\ \Hline
         {\bf Robots}  & 1.78M& 0.26  & 0.46M & 0.056 & 2.0M & 0.085& 0.52M & 0.043 \\
         {\bf ego-Facebook}  & 97.4M& 18.3  & 15.4M  & 2.0  & 116.9M &6.0  & 20.9M & 131.5  \\
         {\bf Youtube} & 27.6G& 57{,}653 & 2.3G & 8{,}116 & 34.0G & 28{,}870 & 8.7G & 7{,}283 \\  
         {\bf Advogato} & 56.7M& 11.9& 20.5M & 8.53 & 74.4M  &3.5  & 29.0M & 2.3\\
         {\bf StringHS} & 1.4G& 521.4  & 1.3G & 335.9 & 6.0G  & 260.4  & 2.8G &237.3\\
         {\bf StringFC} & 1.0G& 852.7& 0.97G & 664.7 & 5.1G & 596.0  & 2.3G & 579.8\\
         {\bf BioGrid}  & 1.7G& 495.1& 0.40G & 86.6& 2.5G & 137.0& 0.63G & 47.9 \\
         {\bf Epinions}  & 3.5G& 849.6 &1.0G  & 229.6  & 4.5G & 264.1 & 1.3G & 122.8\\
         {\bf WebGoogle}  & - & -& 4.7G & 724.5& - & -  & 5.2G & 444.7\\
         {\bf WikiTalk}   & -& -& 14.4G & 3{,}064 & - & -  & 16.0G & 1{,}060\\
         {\bf YAGO}  & - & -&3.8G & 809.4& - & -  & 3.9G & 589.7\\
         {\bf CitPatents}  & - &  -& 2.2G & 469.5& - & -  & 2.5G & 226.4\\
         {\bf Wikidata}  & - & -& 1.44G & 267.5& - & -  & 1.47G & 70.4\\
         {\bf Freebase}  & - & -& 1.0G & 5{,}696& - & -  & 3.7G & 5{,}176\\\hline
         {\bf g-Mark-1m}  & - & -& 0.63G & 104.9 & -  &-& 0.64M & 55.7\\
         {\bf g-Mark-5m} & - & - & 4.1G & 728.0& - &-& 4.1G & 257.4\\
         {\bf g-Mark-10m}  & - &-& 9.3G & 1{,}715& 	-&-& 9.4G & 574.2\\
         {\bf g-Mark-15m} & -  &-& 13.8G & 2{,}741 &-&-& 13.9G & 862.5\\
         {\bf g-Mark-20m} & -& -& 20.3G& 4{,}251 &-&-& 20.6G & 1{,}590\\\hline
    \end{tabular}
    }
\end{table*}

\begin{table*}[!t]
\begin{minipage}[t]{.43\textwidth}
    \centering
    \caption{Update time on \cpqindex}
    \label{tab:maitenance_structural}
    {\small
    \begin{tabular}{|c|r|r|}\hline
         \multirow{2}{*}{Dataset}  & \multicolumn{1}{|c|}{Edge} & \multicolumn{1}{|c|}{Edge} \\ 
         & \multicolumn{1}{|c|}{deletion} & \multicolumn{1}{|c|}{insertion} \\\Hline
         {\bf Robots}   & 0.0008 [s] & 0.0005 [s] \\
         {\bf Advogato} & 0.005 [s] & 0.001 [s] \\
         {\bf BioGrid}  & 0.6 [s] & 0.2 [s] \\         
         {\bf StringHS} & 0.3 [s] & 0.1 [s]\\        
         {\bf StringFC} & 0.2 [s] & 0.06 [s] \\
         {\bf Youtube}  & 0.9 [s] & 0.3 [s]\\\hline
    \end{tabular}
    }
\end{minipage}
  \hfill
  \begin{minipage}[t]{.565\textwidth}
    \centering
    \caption{Update time on \iacpqindex}
    \label{tab:maitenance_wa_structural}
    {\small
    \begin{tabular}{|c|r|r|r|r|}\hline
         \multirow{2}{*}{Dataset}  & \multicolumn{1}{|c|}{Edge} & \multicolumn{1}{|c|}{Edge} & \multicolumn{1}{|c|}{Label sequence} & \multicolumn{1}{|c|}{Label sequence} \\ 
         &\multicolumn{1}{|c|}{deletion}&\multicolumn{1}{|c|}{insertion}&\multicolumn{1}{|c|}{deletion}&\multicolumn{1}{|c|}{insertion}\\
         \Hline
         {\bf Robots}   & 0.0002 [s] & 0.0001 [s]& 0.2 [$\mu$s] & 0.01 [s] \\
         {\bf Advogato} & 0.004 [s] & 0.0004 [s]& 0.3 [$\mu$s] & 0.7 [s] \\
         {\bf BioGrid}  & 1.0 [s] & 0.005 [s]& 0.5 [$\mu$s] & 14.2 [s] \\        
         {\bf StringHS} & 0.2 [s] & 0.03 [s]& 0.5 [$\mu$s] & 15.4 [s]\\        
         {\bf StringFC} & 0.2 [s] & 0.04 [s]& 0.5 [$\mu$s] & 9.8 [s] \\
         {\bf Youtube}  & 1.2 [s] & 1.1 [s]& 0.5 [$\mu$s] & 255.5 [s]\\
         {\bf YAGO}  & 0.7 [s] & 0.04 [s]& 0.8 [$\mu$s] & 30.3 [s]\\
         {\bf Wikidata}  & 0.7 [s] & 0.4 [s]& 1.0 [$\mu$s] & 24.2 [s]\\         
         {\bf Freebase}  & 1.7 [s] & 0.2 [s]& 1.1 [$\mu$s] & 21.8 [s]\\
         \hline
    \end{tabular}
    }
    \end{minipage}
\end{table*}

\subsection{Are CPQ-aware indexes compact?}
\label{ssec:exp_scalability}
We can also give a positive answer to this question.
Table \ref{tab:index} shows the index sizes and index times.
\cpqindex{} achieves smaller index size than Path,
because it stores a single s-t pair regarding to a class while Path stores multiple s-t pairs regarding to label sequences.
\CPQ-aware indexes can reduce its size well when the graph structures and labels have large skews. 
\iacpqindex{} is much smaller than \cpqindex{} because it stores s-t pairs in the given interest.
In WikiTalk,  WebGoogle, Yago, Wikidata, and Freebase, the interest-{\it un}aware indexes cannot be constructed due to their size.
The interest-aware indexes work well for large graphs, where index size is controllable by specifying the appropriate interests. 

Indexing time in both \cpqindex{} and Path are generally large.
Comparing \cpqindex{} with Path, \cpqindex{} is less efficient than Path because \cpqindex{} requires computing $k$-path-bisimulation, while Path just enumerates s-t pairs with label sequences.
These difference is not very large in most datasets.
Comparing the interest-aware indexes with the interest-unaware indexes, the interest-aware indexes are more practical, which clearly takes less time for construction.
\iacpqindex{} contributes to the scalable and efficient index construction with query acceleration. 
In this evaluation, we set $k$ as two. We show that \iacpqindex{} can be constructed in larger $k$ in Sec.~\ref{ssec:exp_analysis}.

\begin{figure}[t]
    \centering
	\includegraphics[width=0.7\linewidth]{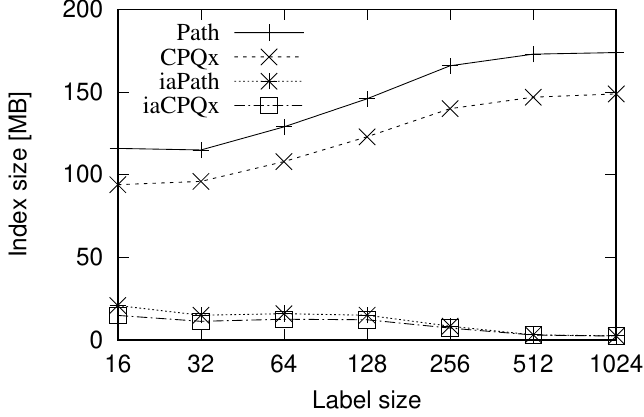}
    \caption{Impact of the size of labels on index size on ego-Facebook}
    \label{fig:labelsize}
    \vspace{-5mm}
\end{figure}

Figure~\ref{fig:labelsize} shows the index size on ego-Facebook varying the size of labels from 16 to 1024.
This results show the index sizes of Path and \cpqindex{} gradually increase because the cardinally of label sequences and the number of class identifier increase, repsectively.
While the index sizes of iaPath and \iacpqindex{} decrease as the size of labels increases.
This is because these interest-aware indexes store only paths that are matched with the given interests.
So, if the number of labels is large, the number of paths that match the interests is small.
Comparing CPQ-aware indexes and language-unaware path indexes, the size of indexes of CPQ-aware indexes are always smaller than those of language-unaware path indexes.
This indicates that our indexes have robustness to the number of labels, in particular, \iacpqindex.

\begin{figure*}[ttt]
\begin{tabular}{ccc}
    \begin{minipage}{.56\textwidth}
    \begin{minipage}[t]{1.0\linewidth}
    \centering
    \includegraphics[width=0.9\linewidth]{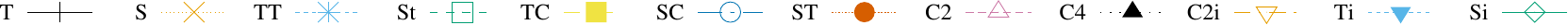}
\end{minipage}
\\
\begin{minipage}[t]{1.0\linewidth}
     \centering
      \includegraphics[width=0.9\linewidth]{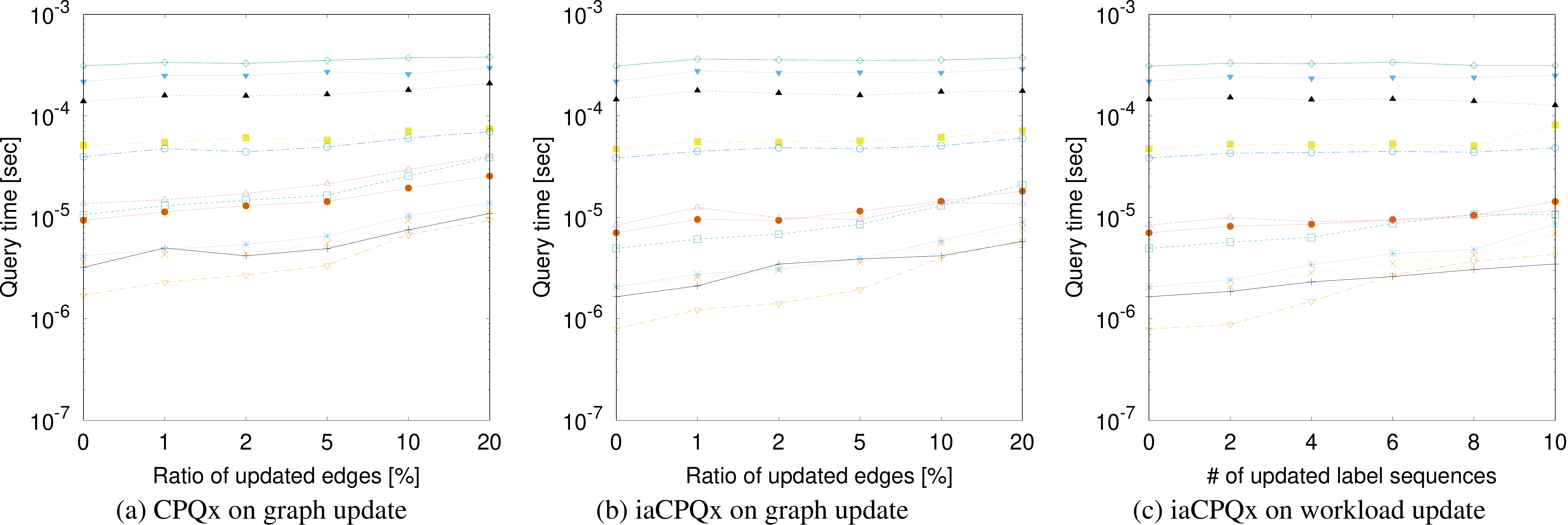}
\end{minipage}
    \caption{Impact of maintenance to query time on Robots}
    \label{fig:maintenance_querytime}
  \end{minipage}
  \begin{minipage}{.40\textwidth}
  \centering
\tblcaption{The increasing ratio of index size on update on Robots
}
\label{tab:maintenance_indexsize}
{\small
\begin{tabular}{|c|rrrrr|}\hline
    \multirow{2}{*}{Index}&\multicolumn{5}{|c|}{Ratio of updated edges}\\
        & 1 & 2 & 5 & 10 & 20 \\\hline
     \cpqindex &1.02& 1.04 & 1.11  &  1.35 & 1.63 \\
     \iacpqindex & 1.03 & 1.06 & 1.13 & 1.31 & 1.53 \\\hline\hline
     \multirow{2}{*}{Index}&\multicolumn{5}{|c|}{\# of updated label sequences}\\
     & 2 & 4 & 6 & 8 & 10 \\\hline
    \iacpqindex & 1.002 & 1.06  &  1.15 & 1.24 & 1.48 \\\hline
\end{tabular}
}
  \end{minipage}

\end{tabular}
\end{figure*}

 \begin{figure*}[!t]
\begin{minipage}[t]{1.0\linewidth}
    \centering
    \includegraphics[width=0.95\linewidth]{figures/experiment/k_query_legends.pdf}
    \end{minipage}
    \\
\begin{minipage}[t]{1.0\linewidth}
    \centering
    \includegraphics[width=1.0\linewidth]{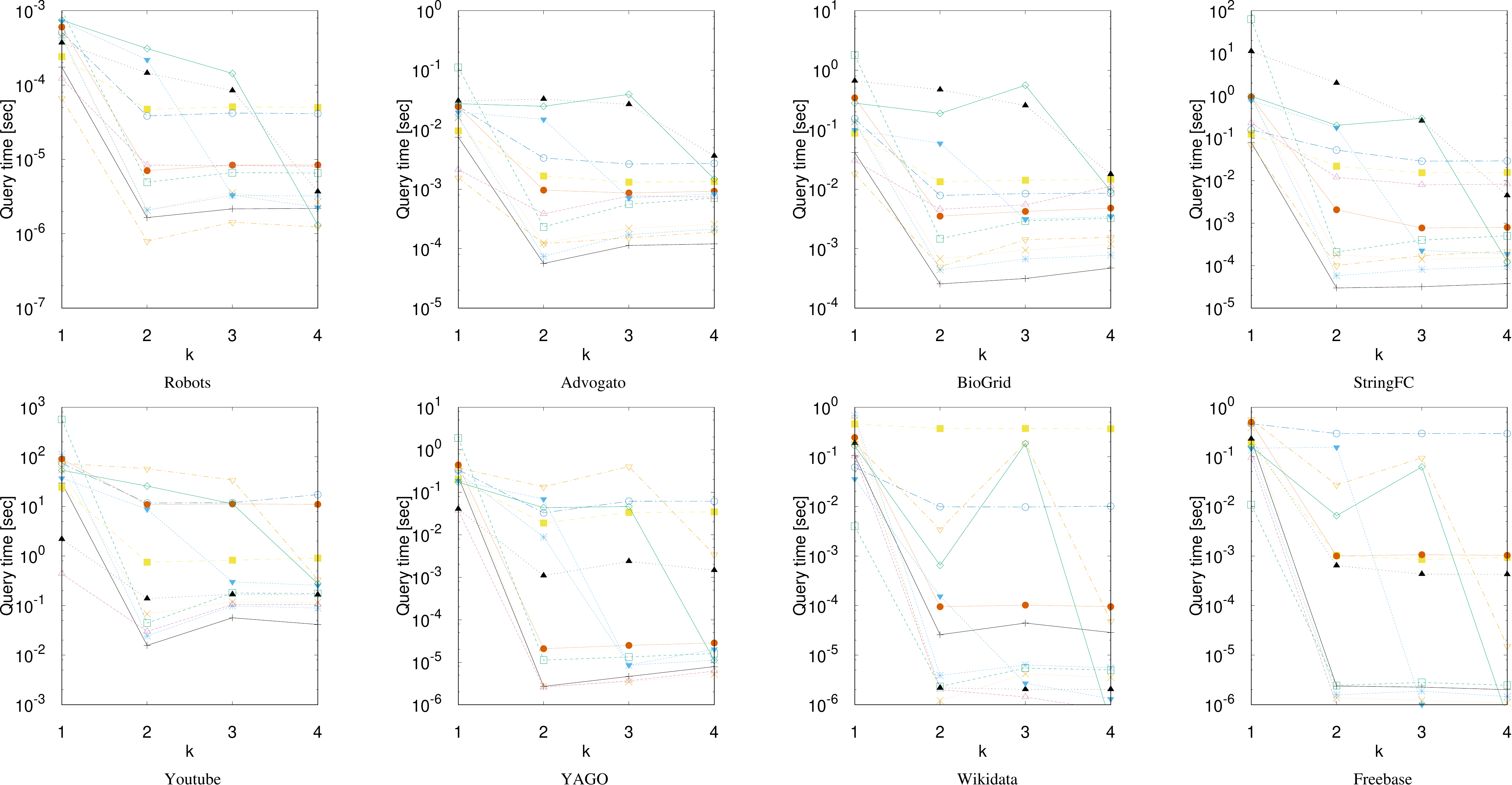}
    \caption{Impact of $k$ on query time with \iacpqindex}
    \label{fig:querytime_wa_structural_index_k}
        \end{minipage}
\end{figure*}

\subsection{Are CPQ-aware indexes maintainable?}
\label{ssec:exp_update}
In short, we can answer this question affirmatively. 
\CPQ-aware indexes can be efficiently updated with small deterioration of query processing and a small increase of its sizes.

\noindent
{\bf Update time.} 
To study the impact of graph and interest updates, 
we delete and insert a hundred edges and label sequences of C2 queries, respectively, and report the average response time of each operation.
Table \ref{tab:maitenance_wa_structural} shows
the update time on \iacpqindex.
Our indexes can be quickly updated compared to the initial index construction time. 
Thus, our indexes can handle graph and interest updates on large graphs.

\noindent
{\bf Impact of updates on query time and index size.}  Our update method lazily updates our index, and thus it deteriorates performance of query time and increases the index size.
We here evaluate the query time and index size after deleting $x\%$ edges (resp. $x$ label sequences) and inserting the deleted edges (resp. label sequences).

Figure \ref{fig:maintenance_querytime} shows the query time after updates.
The query templates whose query times are small (e.g., C2i and T) increase their query time after updates because of increasing {\sc LookUp} costs. 
On the other hand, the query templates with large {\sc Join} costs (e.g., C4 and Si) do not increase their query time much because lookup costs are relatively small compared with {\sc Join} costs.
Note that we confirmed that the query results are the same before and after updates.

Table \ref{tab:maintenance_indexsize} shows the increasing ratio of index size after updates.
Our update method does not merge two class identifiers even if the s-t pairs in blocks with the class identifiers are $k$-path-bisimilar, and thus the size of index increases.
We can see that our update method is practical because the increasing ratio is small even if the number of updates is large.

 \begin{figure*}[!t]
\begin{minipage}[t]{1.0\linewidth}
    \centering
    \includegraphics[width=0.95\linewidth]{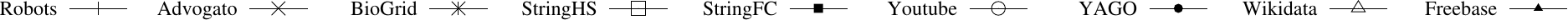}
    \end{minipage}
    \\
\begin{minipage}[t]{1.0\linewidth}
     \centering     \vspace{-5mm}
  \subfloat[Index size]{\epsfig{file=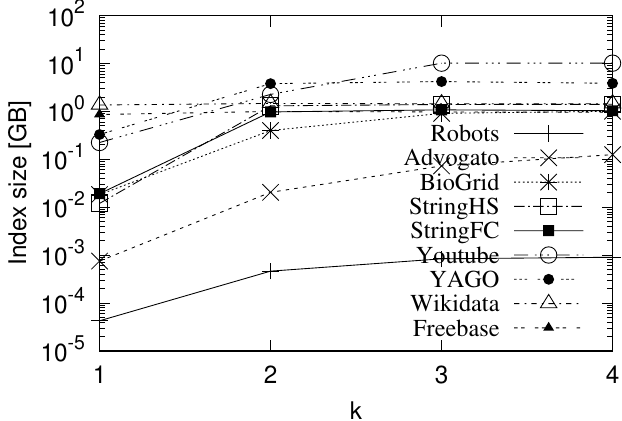,width=0.45\linewidth}
  }
  \subfloat[Index time]{\epsfig{file=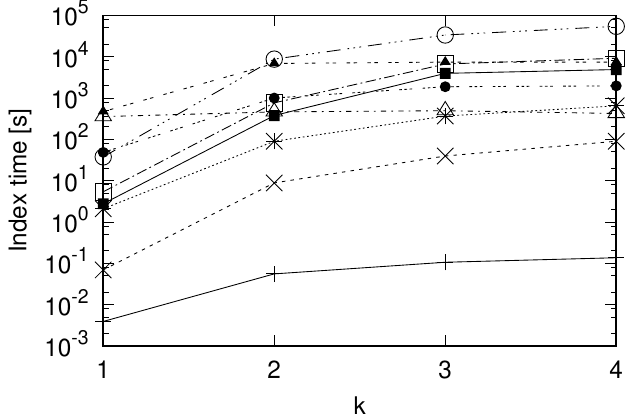,width=0.45\linewidth}
  }
\end{minipage}
    \caption{Impact of $k$ on index construction of \iacpqindex}
    \label{fig:indexconstruction_k}
\end{figure*}

\subsection{Are CPQ-aware indexes well-behaved as $k$ grows?}
\label{ssec:exp_analysis}
We can also give a positive answer.
As $k$ increases, query processing time accelerates substantially.

Figure~\ref{fig:querytime_wa_structural_index_k} shows the query time for \iacpqindex{} varying with $k$.
We can see that the query time decreases from $k=1$ to $k=2$. Some query times increase when $k>2$. There are two reasons for this. First, CPQ-aware indexes divide paths into too fine granularity for some query templates, and which increases {\sc LookUp} and {\sc Conjunction} costs.
Second, costs of {\sc Join} possibly increase, in particular, query time of C4 and Si queries increases when $k=3$. 
As we described in Section~\ref{ssec:indexdef}, queries with diameters $i$ become the fastest when $k = i$. 

Figure~\ref{fig:indexconstruction_k} shows the index size of \iacpqindex{} and index time for \iacpqindex{} construction varying with $k$, respectively.
The index size increases with increasing $k$ generally.
The increase ratio depends on the number of s-t pairs stored in \iacpqindex.
Therefore, the size of \iacpqindex{} often slightly increases from $k=3$ to $k=4$ because the numbers of length 3 paths and length 4 paths are similar.
In particular, the index size in Freebase does not increase much even when $k$ increases.
The index time increases as increasing $k$ index size.

For deciding appropriate $k$, we can generally select the maximum length of interests when building \iacpqindex.
Otherwise, we select $k$ and the interests to control index construction costs.

\section{Concluding Remarks}
\label{sec:conclusion}

We studied language-aware path indexing for evaluation of \CPQ, a fundamental language at the core of contemporary graph query languages.
We proposed new practical indexes and developed algorithms for index construction, maintenance, and query processing to support the full index life cycle.  
We experimentally verified that our methods achieved up to three orders of magnitude acceleration of query processing over the state-of-the-art, while being maintainable and without increasing index size. 

We highlight three research directions.
 (1) In practice, edges and vertices can also carry local data (e.g., user vertices might have their names and dates of birth)~\cite{Bonifati2018}. Study practical extensions to our methods for supporting \cpq{} combined with querying local data.
 (2) Investigate practical methods for scalable index construction that adaptively controls interests and $k$. 
 (3) Now that we have practical \cpq-aware indexes, they can be used in a standard query processing pipeline, i.e., queries expressed in practical languages such as SPARQL and Cypher can use our indexes as part of a physical execution plan. 
 Study query compilation and optimization strategies for \CPQ{} combined with other languages such as {\em RPQ} and {\em CQ}.




\noop{
\clearpage

\subsubsection{Impact of query plans}
The query plans should be affected the query time. In conjunctive path quereis, the query plan corresponds to orders of joins and conjunction operations. 
Here, we evaluate the query time varying orders of joins and conjunctions.

Table \ref{tab:queryplan} shows the averages of the minimum query time and the maximum query time for three queries, \sf{T}, \sf{St}, and \sf{Ti} on StringFC and Youtube datasets \footnote{The reason that we choose these queries are that they only have several query plans when $k$ is two.}.
The difference between min and max query times of Ti query becomes large comparing other queries.
This is because only Ti query includes join operations and the orders of join affects the query time.
On the other hand, other queries do not affect much the query plans.
This result indicates that it is better to use query optimization techniques if queries includes join operations.

\begin{table}[]
    \centering
    \caption{Impact of Query Plans}
    \label{tab:queryplan}
    \begin{tabular}{|c|c|c|c|c|}\hline
         \multirow{2}{*}{Query}& \multicolumn{2}{|c|}{StringsFC} & \multicolumn{2}{|c|}{Youtube} \\
          &\multicolumn{1}{c}{min QT}&\multicolumn{1}{c|}{max QT}&\multicolumn{1}{|c}{min QT}&\multicolumn{1}{c|}{max QT}\\ \Hline
         {\bf T} & 0.23 [ms] & 0.36 [ms] &0.14 [s] & 0.23 [s] \\
         {\bf St} & 1.43 [ms] & 1.55 [ms] & 0.56 [s] & 0.64 [s]\\
         {\bf Ti} & 32.8 [ms] & 177.0 [ms] & 7.12 [s] &  8.2 [s]\\        \hline
    \end{tabular}
\end{table}

\begin{example}
Figure \ref{fig:graph} shows an example of edge labeled graph. 
Figure \ref{fig:bisimulation_example} shows $k$-path bisimulation in Figure \ref{fig:graph} where $k=2$.
The paths within dashed rectangles denote that they are $k$-bisimilar in Figure \ref{fig:bisimulation_example}, for example, ($v_1,v_2$) and ($v_6,v_5$) are 1-path-bisimilar and ($v_2$, $v_6$) and ($v_5$, $v_1$) are 2-path-bisimilar.
\end{example}

\begin{figure*}[ttt]
    \centering
    \scalebox{0.6}{
    \begin{tikzpicture}[->,>=stealth',shorten >=1pt,auto,node distance=2cm,
  thick,main node/.style={circle,draw,font=\sffamily\small}]

        scale only axis,
        width             = 6.5cm,
        enlarge y limits  = {abs=0.5},
        axis y line*      = middle,
        y axis line style = dashed,
        ytick             = \empty,
        axis x line*      = bottom
      ]
  \draw (0,0) rectangle (6.75,16);
  \node[above right] at (0,15.25)  {\LARGE 1-path-bisumulation};
  
  \node[draw,text width=1.75cm,text centered,very thick]at (1.5,14.5){ada $\to$ ada};
  \draw[dashed] (3,14) rectangle (6,15);
  \node[main node] (10)  at (4.55,14.5) {ada};
  
  \node[draw,text width=1.75cm,text centered,very thick]at (1.5,12.5){ada $\to$ tim\\joe $\to$ zoe};
  \draw[dashed] (3,11.5) rectangle (6,13.5);
  \node[main node] (20)  at (5.5,13) {tim};
  \node[main node] (21)  [left of=20] {ada};
  \node[main node] (22)  at (5.5,12) {zoe};
  \node[main node] (23)  [left of=22] {joe};
  
  \node[draw,text width=1.75cm,text centered,very thick]at (1.5,10.25){ tim $\to$ ada\\zoe $\to$ joe};
  \draw[dashed] (3,9) rectangle (6,11);
  \node[main node] (30)  at (5.5,10.5) {tim};
  \node[main node] (31)  [left of=30] {ada};
  \node[main node] (32)  at (5.5,9.5) {zoe};
  \node[main node] (33)  [left of=32] {joe};
  
  \node[draw,text width=1.75cm,text centered,very thick]at (1.5,7.5){tim $\to$ sue\\zoe $\to$ liz};
  \draw[dashed] (3,6.5) rectangle (6,8.5);
  \node[main node] (40)  at (5.5,8) {sue};
  \node[main node] (41)  [left of=40] {tim};
  \node[main node] (42)  at (5.5,7) {liz};
  \node[main node] (43)  [left of=42] {zoe};
  
  \node[draw,text width=1.75cm,text centered,very thick]at (1.5,5){ sue $\to$ tim\\liz $\to$ zoe};
  \draw[dashed] (3,4) rectangle (6,6);
  \node[main node] (50)  at (5.5,5.5) {sue};
  \node[main node] (51)  [left of=50] {tim};
  \node[main node] (52)  at (5.5,4.5) {liz};
  \node[main node] (53)  [left of=52] {zoe};
  
  \node[draw,text width=1.75cm,text centered,very thick]at (1.5,2){ liz $\to$ sue\\sue $\to$ liz};
  \draw[dashed] (3,0.5) rectangle (6,3.5);
  \node[main node] (60)  at (5.5,3) {sue};
  \node[main node] (61)  [left of=60] {sue};
  \node[main node] (62)  [below of=60] {liz};
  \node[main node] (63)  [below of=61] {liz};

  \draw (6.9,0) rectangle (28,16);
  \node[above right] at (6.9,15.25)  {\LARGE 2-path-bisumulation};
  
  \node[draw,text width=1.75cm,text centered,very thick]at (8.5,14.5){ada $\to$ ada};
  \draw[dashed] (10,14) rectangle (16,15);
  \node[main node] (70)  at (12,14.5) {ada};
  \node[main node] (71)  at (13.25,14.5) {ada};
  \node[main node] (72)  [right of=71] {tim};
  
  \node[draw,text width=1.75cm,text centered,very thick]at (8.5,12.75){ada $\to$ tim};
  \draw[dashed] (10,12.25) rectangle (16,13.25);
  \node[main node] (80)  at (14,12.75) {tim};
  \node[main node] (81)  [left of=80] {ada};

  \node[draw,text width=1.75cm,text centered,very thick]at (8.5,10.5){ada $\to$ sue\\joe $\to$ liz};
  \draw[dashed] (10,9.5) rectangle (16,11.5);
  \node[main node] (90)  at (13,11) {tim};
  \node[main node] (91)  [left of=90] {ada};
  \node[main node] (92)  at (13,10) {zoe};
  \node[main node] (93)  [left of=92] {joe};
  \node[main node] (94)  [right of=90] {sue};
  \node[main node] (95)  [right of=92] {liz};
  
  \node[draw,text width=1.75cm,text centered,very thick]at (8.5,8.25){ tim $\to$ ada};
  \draw[dashed] (10,7.75) rectangle (16,8.75);
  \node[main node] (100)  at (14,8.25) {tim};
  \node[main node] (101)  [left of=100] {ada};
  
  \node[draw,text width=1.75cm,text centered,very thick]at (8.5,6){ tim $\to$ tim\\zoe $\to$ zoe};
  \draw[dashed] (10,5) rectangle (16,7);
  \node[main node] (110)  at (12.5,5.5) {zoe};
  \node[main node] (111)  [left of=110] {joe};
  \node[main node] (112)  at (13.5,5.5) {zoe};
  \node[main node] (113)  [right of=112] {liz};
  \node[main node] (114)  at (12.5,6.5) {tim};
  \node[main node] (115)  [left of=114] {ada};
  \node[main node] (116)  at (13.5,6.5) {tim};
  \node[main node] (117)  [right of=116] {sue};
  
  \node[draw,text width=1.75cm,text centered,very thick]at (8.5,2.75){ tim $\to$ liz\\zoe $\to$ sue};
  \draw[dashed] (10,1.25) rectangle (16,4.25);
  \node[main node] (120)  at (12.5,3.75) {sue};
  \node[main node] (121)  [left of=120] {tim};
  \node[main node] (122)  [below of=120] {liz};
  \node[main node] (123)  at (15.5,1.75) {liz};
  \node[main node] (124)  [left of=123] {zoe};
  \node[main node] (125)  [above of=123] {sue};

  
  \node[draw,text width=1.75cm,text centered,very thick]at (18.75,14){ sue $\to$ ada\\liz $\to$ joe};
  \draw[dashed] (20.25,13) rectangle (26.25,15);
  \node[main node] (130)  at (23.25,13.5) {tim};
  \node[main node] (131)  [left of=130] {ada};
  \node[main node] (132)  at (23.25,14.5) {zoe};
  \node[main node] (133)  [left of=132] {joe};
  \node[main node] (134)  [right of=130] {sue};
  \node[main node] (135)  [right of=132] {liz};
  
  \node[draw,text width=1.75cm,text centered,very thick]at (18.75,10.75){ sue $\to$ sue\\liz $\to$ liz};
  \draw[dashed] (20.25,9.25) rectangle (26.25,12.25);
  \node[main node] (140)  at (24.25,11.75) {sue};
  \node[main node] (141)  [left of=140] {sue};
  \node[main node] (142)  [below of=140] {liz};
  \node[main node] (143)  [below of=141] {liz};
  
  \node[draw,text width=1.75cm,text centered,very thick]at (18.75,7){ sue $\to$ zoe\\liz $\to$ tim};
  \draw[dashed] (20.25,5.5) rectangle (26.25,8.5);
  \node[main node] (150)  at (22.75,8) {sue};
  \node[main node] (151)  [left of=150] {tim};
  \node[main node] (152)  [below of=150] {liz};
  \node[main node] (153)  at (25.75,6) {liz};
  \node[main node] (154)  [left of=153] {zoe};
  \node[main node] (155)  [above of=153] {joe};
  
  \node[draw,text width=1.75cm,text centered,very thick]at (18.75,4.25){ joe $\to$ joe};
  \draw[dashed] (20.25,3.75) rectangle (26.25,4.75);
  \node[main node] (160)  at (24.25,4.25) {zoe};
  \node[main node] (161)  [left of=160] {joe};
  
  \path[every node/.style={font=\small}]
       (10) edge [loop left]  node[above] {} (10)
       
       (20) edge [bend right] node[left] {} (21)
       (21) edge [bend right] node[left] {} (20)
       (22) edge [bend right] node[left] {} (23)
       (23) edge [bend right] node[left] {} (22)
       
       (30) edge [bend right] node[left] {} (31)
       (31) edge [bend right] node[left] {} (30)
       (32) edge [bend right] node[left] {} (33)
       (33) edge [bend right] node[left] {} (32)

       (41) edge [left] node[left] {} (40)
       (43) edge [left] node[left] {} (42)
       
       (51) edge [left] node[left] {} (50)
       (53) edge [left] node[left] {} (52)
       
       (63) edge [bend right] node[left] {} (61)
       (62) edge [bend right] node[left] {} (60)
       (60) edge [bend right] node[left] {} (62)
       (61) edge [bend right] node[left] {} (63)
       
       
       (71) edge [bend right] node[left] {} (72)
       (72) edge [bend right] node[left] {} (71)
       (70) edge [loop left]  node[above] {} (70)
       
       (81) edge [bend right] node[left] {} (80)
       (80) edge [bend right] node[left] {} (81)
       (81) edge [loop left]  node[above] {} (81)
       
       (91) edge [bend right] node[left] {} (90)
       (90) edge [bend right] node[left] {} (91)
       (92) edge [bend right] node[left] {} (93)
       (93) edge [bend right] node[left] {} (92)
       (90) edge [left] node[left] {} (94)
       (92) edge [left] node[left] {} (95)
       
       (101) edge [bend right] node[left] {} (100)
       (100) edge [bend right] node[left] {} (101)
       (101) edge [loop left]  node[above] {} (101)
       
       (111) edge [bend right] node[left] {} (110)
       (110) edge [bend right] node[left] {} (111)
       (112) edge [left] node[left] {} (113)
       (115) edge [bend right] node[left] {} (114)
       (114) edge [bend right] node[left] {} (115)
       (116) edge [left] node[left] {} (117)
       
       (121) edge [left] node[left] {} (120)
       (122) edge [bend right] node[left] {} (120)
       (120) edge [bend right] node[left] {} (122)
       (124) edge [left] node[left] {} (123)
       (123) edge [bend right] node[left] {} (125)
       (125) edge [bend right] node[left] {} (123)
       
       (131) edge [bend right] node[left] {} (130)
       (130) edge [bend right] node[left] {} (131)
       (132) edge [bend right] node[left] {} (133)
       (133) edge [bend right] node[left] {} (132)
       (130) edge [left] node[left] {} (134)
       (132) edge [left] node[left] {} (135)
       
       (143) edge [bend right] node[left] {} (141)
       (142) edge [bend right] node[left] {} (140)
       (140) edge [bend right] node[left] {} (142)
       (141) edge [bend right] node[left] {} (143)
       
       (151) edge [left] node[left] {} (150)
       (152) edge [bend right] node[left] {} (150)
       (150) edge [bend right] node[left] {} (152)
       (154) edge [left] node[left] {} (153)
       (153) edge [bend right] node[left] {} (155)
       (155) edge [bend right] node[left] {} (153)
       
       (161) edge [bend right] node[left] {} (160)
       (160) edge [bend right] node[left] {} (161);
\end{tikzpicture}
}
    \caption{1-path-bisimuation and 2-path-bisimulation in Figure \ref{fig:graph}}
    \label{fig:bisimulation_example_appendix}
\end{figure*}

\begin{figure*}[ttt]
\begin{minipage}[t]{1.0\linewidth}
	\centering
	\includegraphics[width=0.95\linewidth]{figures/k_query_legends-crop.pdf}
	\end{minipage}
	\\
\begin{minipage}[t]{1.0\linewidth}
	\centering
	\includegraphics[width=0.95\linewidth]{figures/k_query_path-crop.pdf}
	\caption{Query time of path index varying $k$}
	\label{fig:querytime_path_index_k}
		\end{minipage}
\end{figure*}

\subsubsection{A breakdown of index size}

Our structural index has two entries.
We here show that the ratio of sizes of entries in the structural index.

Table \ref{tab:breakdown} shows the size of entries for all datasets.
The size of $I_{h2p}$ is generally larger than that of $I_{l2p}$ because history identifiers are stored multiple time while paths are stored a single time.
In the structural index of {\bf Youtube}, the size of $I_{l2p}$ is larger than that of $I_{h2p}$.
This is because paths do not often become $k$-path-bisimilar since {\bf Youtube} has large degrees.

\begin{table}[ttt]
    \centering
    \caption{A breakdown of index size.}
    \label{tab:breakdown}
    \begin{tabular}{|c|r|r|r|r|}\hline
         \multirow{2}{*}{Dataset}& \multicolumn{2}{|c|}{Structural}&\multicolumn{2}{|c|}{WA-structural} \\
          &\multicolumn{1}{c}{$I_{h2p}$ }&\multicolumn{1}{c|}{$I_{l2h}$}&\multicolumn{1}{c}{$I_{h2p}$}&\multicolumn{1}{c|}{$I_{l2h}$}\\ \Hline
         {\bf Robots}  & 1.62 & 0.16 [MB]& 0.4& 0.02 [MB] \\
         {\bf Advogato} & 37.4 & 19.3 [MB] & 19.0 & 0.9 [MB] \\
         {\bf BioGrid}  & 1.07& 0.59 [GB] & 0.35 & 0.03 [GB]\\
         {\bf StringHS} & 1.32 & 0.11 [GB] & 1.3 & 0.06 [GB] \\
         {\bf StringFC} & 0.98 & 0.66 [GB]&0.9& 0.02 [GB] \\
         {\bf Youtube} & 3.25 & 24.2 [GB]& 1.7& 0.5 [GB] \\\hline
    \end{tabular}
    \end{table}

\subsubsection{The numbers of history identifiers and paths}
The structural index achieves faster query processing for conjunction operators.
This is because our query processing algorithm compares history identifiers instead of paths.
We here show the numbers of history identifiers and paths for square queries.

Table \ref{tab:numhistory} shows the number of history identifiers in structural and WA-structural index and the number of paths in path index.
The numbers of history identifiers and paths have large gaps, in particular, StringHS and StringFC.
Thus, structural index achieves much faster query time of queries with conjunction in StringHS and StringFC (see Fig. \ref{fig:querytime}).

\begin{table}[]
    \centering
    \caption{The difference between the numbers of history identifiers and paths}
    \label{tab:numhistory}
    \begin{tabular}{|c|r|r|r|}\hline
          \multicolumn{1}{|c|}{Dataset}&\multicolumn{1}{|c|}{Structural} &\multicolumn{1}{|c}{WA-Structural}&\multicolumn{1}{|c|}{Path}\\ \Hline
         {\bf Robots} & 0.5 K & 0.17 K  & 3.0 K \\
         {\bf Advogato} & 50 K & 7.1 K & 150 K \\
         {\bf BioGrid} & 405 K  & 82 K  & 970 K \\         
         {\bf StringHS} & 101 K & 4.9 K&  18 M \\        
         {\bf StringFC} & 61 K & 5.5 K & 4.1 M\\
         {\bf Youtube} & 22 M & 2.1 M & 27 M\\         \hline
    \end{tabular}
\end{table}

\subsubsection*{Algorithm} Algorithms \ref{alg:deleteedge} and \ref{alg:addedge} show pseudo-code of edge deletion and edge insertion, respectively.
These algorithms share the same procedures; enumerating $k$ paths, checking alternative paths, and finding history identifiers (lines 2 -- 12).
The difference between them is updating $I_{h2p}$; in edge deletion, $I_{h2p}$ is added if deleted paths have other label sequences from deleted ones, and in edge insertion, $I_{h2p}$ is always updated because paths are definitely added new label sequences.
\begin{algorithm}[!t] 
	\caption{Update for inserting edge}	\label{alg:addedge}
		\DontPrintSemicolon
			    \SetKwInOut{Input}{input}
	            \SetKwInOut{Output}{output}
	            \SetKwFunction{Expand}{Expand}
	            \Input{Index $I_{[\graph]_k}$, \graph, inserted edge $e_i = (v,v',\elabel)$}
	            \Output{Updated Index}
            	{\bf procedure} {\sc EdgeInsertion}\\
            	$\spaths \leftarrow$ Enumerate paths at most length $k$ involved $e_i$;\\
            	\For{$(v,u) \in \spaths$}{
                	\If{there are alternative paths of $(v,u)$}{
                	    Remove $(v,u)$ from $\spaths$;
                	}
            	}
            	\For{$(v,u) \in \spaths$}{
            	    updateflag$\leftarrow$ false;\\
                	\For{$h \in \histories$}{
                	    \If{$(v,u) \in I_{h2p}(h)$}{
                	    delete $(v,u)$ from $I_{h2p}(h)$;\\
                	    $h_d \leftarrow h$;\\
                	    updateflag$\leftarrow$ true;\\
                	    {\bf break};\\
                	    }
                	}
                	$h_{new}\leftarrow$ new history identifier;\\ 
                	$I_{h2p}$.attend($h_{new},(v,u)$);\\
                	\If{updateflag}{
                     	\For{$\overline{\elabel} \in \elabels^k$}{
                         	\For{$h \in I_{l2h}(\overline{\elabel})$}{
                         	    \If{$h = h_d$}{
                         	        $I_{l2h}$.append($\overline{\elabel}, h_{new}$);
                         	    }
                     	    }
                     	}
                 	}
              }
              {\bf return} $I_{k}$;\\
              {\bf end procedure}
\end{algorithm}

\begin{figure}[ttt]
  \centering
   \subfloat[Structural]{\epsfig{file=./figures/k_indextime_bisimulation.eps,width=0.5\linewidth}
   }
    \subfloat[WA-Structural]{\epsfig{file=./figures/k_indextime_workload.eps,width=0.5\linewidth}
     }   
   \caption{Impact of $k$ for index time}
   \label{fig:indextime_k}
\end{figure}
}

\section*{Acknowledgement}
This work was supported by JST PRESTO Grant Number JPMJPR21C5 and JSPS KAKENHI Grant Number JP20H00583, Japan.

\bibliographystyle{IEEEtranS}
\bibliography{vldb2020_cpq}

\end{document}